\documentclass{article}
\usepackage{authblk}
\usepackage[utf8]{inputenc}
\usepackage{geometry}

\usepackage{amsmath}
\usepackage{graphicx}
\usepackage[colorlinks=true, allcolors=blue]{hyperref}
\usepackage{color}
\usepackage{graphicx,epsfig}
\usepackage{amsfonts,amssymb,amsmath}
\usepackage{longtable}
\usepackage{float}
\usepackage{xcolor}
\usepackage{color}
\usepackage{multirow}
\usepackage{array}
\usepackage{wrapfig}

\newgeometry{vmargin={25mm}, hmargin={25mm,25mm}}

\newcommand{\cevns} {CE$\nu$NS}

\title{Theory of Neutrino Physics --Snowmass TF11 (aka NF08) Topical Group Report}

\author[1]{Andr\'e de Gouv\^ea\footnote{Topical Group Convener; degouvea@northwestern.edu}}
\author[2]{Irina Mocioiu\footnote{Topical Group Convener; ium4@psu.edu}}
\author[3]{Saori Pastore\footnote{Topical Group Convener; saori@wustl.edu}} 
\author[4]{Louis E. Strigari\footnote{Topical Group Convener; strigari@tamu.edu}}
\author[5]{L.~Alvarez-Ruso}
\author[6]{A.~M.~Ankowski}
\author[7]{A.B.~Balantekin}
\author[1,8]{V.~Brdar}
\author[9]{M.~Cadeddu}
\author[10]{S.~Carey}
\author[11]{J.~Carlson}
\author[12]{M.-C.~Chen}
\author[13]{V.~Cirigliano}
\author[13]{W.~Dekens}
\author[14]{P.~B.~Denton}
\author[15]{R.~Dharmapalan}
\author[7]{L.~Everett}
\author[16]{H.~Gallagher}
\author[8]{S.~Gardiner}
\author[14]{J.~Gehrlein}
\author[17,18]{L.~Graf}
\author[17,19]{W.C.~Haxton}
\author[20]{O.~Hen}
\author[21]{H.~Hergert}
\author[22]{S.~Horiuchi}
\author[23]{P~.Q.~Hung}
\author[24]{J.~Isaacson}
\author[25]{N.~Jachowicz}
\author[26]{L.~Jin}
\author[27]{A.N.~Khan}
\author[28]{A.~Lovato}
\author[8]{P.~A.~N.~Machado}
\author[29]{K.~Mahn}
\author[15]{D.~Marfatia}
\author[22]{C.~Mariani}
\author[11]{E.~Mereghetti}
\author[24]{J.G.~Morf\'{\i}n}
\author[30]{A.~Nicholson}
\author[10]{G.~Paz}
\author[31]{R.~Plestid}
\author[24]{N.~Rocco} 
\author[32]{I.~Sarcevic}
\author[33]{R.~Schiavilla}
\author[34]{A.~Sousa}
\author[35]{J.~Tena-Vidal}
\author[24]{M.L.~Wagman}
\author[36]{A.~Walker-Loud}
\affil[1]{Physics \& Astronomy Department, Northwestern University, Evanston, IL}
\affil[2]{Department of Physics, The Pennsylvania State University, University Park, PA}
\affil[3]{Department of Physics and the McDonell Center for the Space Sciences, Washington University
in Saint Louis, Saint Louis, MO}
\affil[4]{Department of Physics and Astronomy, Mitchell Institute for Fundamental Physics and Astronomy, College Station, TX}
\affil[5]{Instituto de Física Corpuscular (IFIC), Consejo Superior de Investigaciones Científicas (CSIC) and Universidad de Valencia, Spain}
\affil[6]{SLAC National Accelerator Laboratory, Stanford University, Menlo Park, CA}
\affil[7]{University of Wisconsin, Madison; Department of Physics; Madison, WI}
\affil[8]{Particle Theory Department, Fermi National Accelerator Laboratory, Batavia, IL}
\affil[9]{Istituto Nazionale di Fisica Nucleare (INFN), Sezione di Cagliari, Complesso Universitario di Monserrato - S.P. per Sestu Km 0.700, 09042 Monserrato (Cagliari), Italy}
\affil[10]{Department of Physics and Astronomy, Wayne State University, Detroit, Michigan}
\affil[11]{Los Alamos National Laboratory, Los Alamos, NM}
\affil[12]{Department of Physics and Astronomy, University of California, Irvine, CA}
\affil[13]{Institute for Nuclear Theory, University of Washington, Seattle WA}
\affil[14]{High Energy Theory Group, Physics Department, Brookhaven National Laboratory, Upton, NY}
\affil[15]{Department of Physics and Astronomy, University of Hawaii at Manoa, Honolulu, HI}
\affil[16]{Department of Physics and Astronomy, Tufts University, Medford, MA}
\affil[17]{Department of Physics, University of California, Berkeley, CA}
\affil[18]{Department of Physics, University of California, San Diego, CA}
\affil[19]{Lawrence Berkeley National Laboratory, Berkeley, CA}
\affil[20]{Massachusetts Institute of Technology, Cambridge MA}
\affil[21]{Facility for Rare Isotope Beams and Department of Physics \& Astronomy, Michigan State University, East Lansing, MI}
\affil[22]{Center for Neutrino Physics, Department of Physics, Virginia Tech, Blacksburg, VA}
\affil[23]{Department of Physics, University of Virginia, Charlottesville, VA}
\affil[24]{Fermi National Accelerator Laboratory, Batavia, IL}
\affil[25]{Ghent University, Department of Physics and Astronomy, Gent, Belgium}
\affil[26]{Department of Physics, University of Connecticut, Storrs, CT}
\affil[27]{Max-Planck-Institut fur Kernphysik Postfach 103980 69029 Heidelberg}
\affil[28]{Physics Division, Argonne National Laboratory, Argonne, IL}
\affil[29]{Department of Physics \& Astronomy, Michigan State University, East Lansing, MI}
\affil[30]{Department of Physics and Astronomy, University of North Carolina, Chapel Hill, NC}
\affil[31]{Walter Burke Institute for Theoretical Physics, California Institute of Technology, Pasadena, CA}
\affil[32]{Department of Physics and Department of Astronomy, University of Arizona, Tucson, AZ}
\affil[33]{Physics Department, Old Dominion University, Norfolk VA 23529 and Theory Center, Jefferson Lab, Newport News, VA}
\affil[34]{Department of Physics, University of Cincinnati, Cincinnati, OH}
\affil[35]{Tel Aviv University, Tel Aviv 69978, Israel}
\affil[36]{Nuclear Science Division, Lawrence Berkeley National Laboratory}

\date{\today}

\begin{document}

\maketitle

\section{Executive Summary (and Introduction)} 

The discovery of nonzero neutrino masses requires new fundamental fields and new interactions. We know very little about this new physics other than the fact that it exists. The new degrees of freedom can be fermions or bosons, ultra-light or super-heavy, charged or neutral, within reach of experimental efforts being pursued today or virtually invisible to any foreseeable future experiment. At the same time, the discovery of a new mixing matrix -- very different from the quark one -- serves as a new, perhaps decisive, piece to the elusive flavor puzzle.  

Experimentally, the path forward is well defined: a diverse neutrino program is required in order to explore the new physics revealed in the neutrino sector. It includes very intense neutrino beams, very large, finely instrumented detectors, very large, ultra-clean detectors to search for neutrinoless double-beta decay, and novel detectors for precision measurements of beta-decay. In the next decade, a deluge of neutrino-related data is expected. 

A robust neutrino theory and phenomenology effort is required in order to exploit these unique probes of fundamental physics, interpret the data, build models to accommodate new phenomena, provide guidance for current and future experimental efforts, and connect the new discoveries in neutrino physics to other areas of particle and nuclear physics, astrophysics, and cosmology. Here we highlight the role and the goals of neutrino theory, concentrating on how it complements and contributes to theoretical efforts in other areas of fundamental physics, and on some of the challenges for the near future and the coming decades. The main message is that the theoretical neutrino effort is both very broad in terms of the tools required and rather focused  when it comes to individual physics challenges. In summary, neutrino theory requires a broad set of tools in order to attack a unique set of physics problems. 

Among the goals of neutrino theory is to identify the different hypothetical degrees of freedom and interactions responsible for nonzero neutrino masses. More progress requires a coherent theoretical and phenomenological effort to establish connections to other outstanding questions in fundamental particle physics ranging from quantum gravity to the mechanism of baryogenesis to the dark matter puzzle. On the phenomenology side, the physics behind nonzero neutrino masses can manifest itself in, to name a few, neutrino oscillations, fundamental electric-dipole moments, the $g-2$ of charged fermions, charged-lepton flavor violating processes, or high energy colliders. Theory is required in order to explore these connections and identify promising new directions.   

The flavor puzzle -- understanding the underlying fundamental physics that is responsible for the patterns observed in the quark and lepton masses and mixing parameters -- is also the subject of theoretical physics research. It may prove to be one way in which ingredients of a more fundamental theory of nature, including string theory, manifest themselves, and it may contain information associated to grand unification. On the other hand, the fact we don't yet have all the pieces of the lepton mixing matrix in place allows one to test different general principles that may lurk behind lepton masses and mixing. Expectations from theories of flavor help provide guidance regarding how well we should measure mixing and other fundamental parameters. 

Long-baseline neutrino oscillations are entering the precision era with the Deep Underground Neutrino Experiment (DUNE) in the United States and Hyper-Kamiokande in Japan. In order to fully exploit these experiments, our quantitative understanding of neutrino scattering cross sections must improve significantly across a wide range of energies where different reaction mechanism are at play, e.g., quasi-elastic scattering, resonance-dominated scattering, shallow and deep inelastic scattering. Furthermore, the fact that neutrinos only interact weakly, connected with the fact that neutrino oscillation parameters are such that one is forced to work with neutrino energies below a few GeV -- the Earth is not a very large planet --  implies that one needs to understand the scattering cross section of neutrinos off complex nuclei, including carbon, oxygen, and argon. The development of sophisticated simulation tools (event generators) is required to allow for a direct comparison of theoretical calculations and experimental data. Ensuring experiments have the most advanced predictions requires more involvement of theorists in the development and support of generators. Theoretical nuclear physics, along with many-body computational methods, is required in order to accurately describe the targets and the medium through which the products of the collision propagate. A precise description of relatively low-energy neutrino--nucleon scattering depends on the tools of lattice gauge theory. These in turn inform many-body nuclear methods via effective theories based on nucleonic degrees of freedom. Moreover, both LQCD and effective theories provide input and benchmark for more phenomenological models that can cover the broad kinematic range and variety of processes met in experiments. At lower energies, the physics of coherent neutrino--nucleus scattering, first observed a few years ago, needs input from theoretical nuclear physics. On the other end of the energy spectrum, in order to fully exploit the physics of ultrahigh energy cosmic neutrinos, a solid theoretical understanding of neutrino deep-inelastic scattering is required. The energy region in which the degrees of freedom switch from nucleons and pions to partons, referred to as the shallow-inelastic region, requires significant effort from nuclear theorists, lattice theorists, and particle theorists with expertise in perturbative QCD to meet the precision needs of DUNE.

On the more phenomenological side, neutrino theorists explore, very broadly, the physics potential of long-baseline neutrino oscillation experiments. Phenomenological work has revealed, in the last few years, that experimental setups aimed at precisely measuring oscillations can be used to look for relatively light, new degrees of freedom, including candidates for the dark matter and new fermions that may have something to do with nonzero neutrino masses. This type of work serves as motivation for exploring different options for near-detector complexes and inform decisions regarding promising new detector technologies. Phenomenological and model-building efforts are also required in order to interpret neutrino oscillation data, including current and future anomalies. In particle physics experiments, in order to fully exploit experimental data, a coherent experimental-phenomenological-model-building effort is required. Neutrino experiments in general, and neutrino oscillation experiments in particular, are no exception. 
 
Searches for the violation of lepton number inform fundamental neutrino physics and are motivated by neutrino theory in different ways. Different theoretical efforts are required in order to connect lepton-number-violating observables to neutrino properties and other new physics, and in order to connect different lepton-number-violating phenomena to one another. The deepest probe of the violation of lepton number is the search for neutrinoless double-beta decay ($0\nu\beta\beta$). The relation between the lifetime of $0\nu\beta\beta$ and fundamental physics parameters requires precise inputs from theoretical nuclear physics and nucleon physics. Current efforts combine state-of-the-art tools from lattice gauge theory, capable of estimating nucleon-level processes, with those of nuclear physics, that define the nontrivial state of these nucleons inside of the complex nuclei of interest.

Neutrinos are also produced in intense astrophysics environments and, together with photons and gravitational waves, serve as messengers between us and the cosmos. Thanks to the weak-interactions nature of neutrino scattering and their very long lifetimes, these neutrinos can and have been detected on Earth, allowing, with nontrivial input from theory, one to learn more about the different astrophysical processes and the properties of the neutrinos themselves. Type-II supernova explosions in particular are bone fide neutrino factories and the detection of these neutrinos carries invaluable information about these violent phenomena. Intense research in theoretical physics and astrophysics is mandatory in order to interpret what these supernova neutrinos are saying. Assuming neutrino-flavor information is also available, almost guaranteed given the technology employed by detectors capable of seeing supernova neutrinos, one needs to understand the very challenging physics of flavor-transport inside the explosion. In the last couple of decades, significant progress has been made but the correct solution remains elusive given the formidable technical challenges involved.   

At higher energies, neutrinos from the cosmos have been detected with energies between 1~TeV and 10~PeV. These are expected to shed light, with necessary help from astrophysics and astro-particle theory, on the cosmic ray puzzle and the ultimate particle accelerators in the universe. Ultra-high energy neutrinos are also unique probes of new phenomena and have been explored extensively by the phenomenology community. Their very long travel distances and extreme energies are sensitive to neutrino properties, including the neutrino lifetime, and, assuming one can distinguish neutrino flavors, allow for powerful tests of Lorentz invariance and the unitary evolution of quantum mechanical states. 

At the opposite end of the energy spectrum lie the cosmic neutrinos produced at big bang. These have never been directly observed but influence the evolution of the Universe, including the formation of structure. Their existence has been robustly inferred through a variety of different probes, including measurements of the primordial abundances of light nuclei (deuterium, helium, lithium) and precision measurements of the properties of the cosmic microwave background. Theory and phenomenology are required in order to compute the impact of primordial neutrinos and extract, from diverse measurements of the large-scale structure of the Universe, neutrino properties. Some of these translate into nontrivial theoretical and computational physics problems involving many-body physics, nonlinear dynamics, etc. New neutrino properties may be responsible for some of the observed discrepancies in cosmological data, including the $H_0$ puzzle. The exploration of these new properties -- how they impact cosmic surveys, how they can be constrained by Earth-bound experiments -- is the job of particle and astroparticle theorists and cosmologists. 
  
In the last decade, the neutrino theory effort has grown both here and in the rest of the world, but so has the need for a more robust domestic theoretical physics community. Some positive developments include a few dedicated efforts from the DOE and the NSF, increased investment in neutrino theory in the National Labs and a more robust neutrino footprint in the lattice and nuclear physics communities. These are necessary but not sufficient; the domestic neutrino theory community is still very small. To meet the proposed experimental schedules and ambitions, such a qualitative increase in the neutrino theory effort is needed now along with increased support to implement models into and maintain neutrino event generators. Support for theoretical efforts on neutrino scattering at the interface of high-energy and nuclear physics will be critical for achieving reliable cross-section predictions across the range of energies relevant to current and planned neutrino experimental program. A dedicated and coherent effort, with significant investment from the funding agencies, enthusiastic commitment and leadership from the present neutrino theory community, and the support of the particle theory and neutrino experimental communities is absolutely necessary.   
  
In the more detailed discussions below, references are restricted, as much as possible, to the dozens of relevant White Papers submitted by the community during the last two years. These informed most of the content presented here and have been flagged with `WP.'    
  
\section{Neutrino mass and flavor model-building}

The Standard Model of particle physics (SM) can be defined, in a nutshell, as follows. It is a relativistic quantum field theory invariant under $SU(3)_c\times SU(2)_L\times U(1)_Y$ gauge transformations ($c$ for color, $Y$ for hypercharge, $L$ as a reminder of the left-chiral nature of the weak interactions), spontaneously broken to $SU(3)_c\times U(1)_{em}$ ($em$ for electromagnetism). Data are consistent with the hypothesis that spontaneous symmetry breaking is captured by the vacuum expectation value (vev) of the Higgs field $H(1,2,1/2)$, where the numbers in parenthesis correspond to the quantum numbers of the field relative to the $SU(3)_c, SU(2)_L, U(1)_Y$ gauge transformations. The SM also contains several chiral fermion fields, the matter fields: $Q(3,2,1/6)$, $u^c(\bar{3},1,-2/3)$, $d^c(\bar{3},1,1/3)$, $L(1,2,-1/2)$, $e^c(1,1,1)$, all expressed as left-handed Weyl fermions. There are three copies (families or flavors) of every matter field. Neutrinos are the ``top'' component of the lepton-doublet $L$ fields.

Restricting the SM Lagrangian to renormalizable interactions, after elecroweak symmetry breaking, all fermions acquire nonzero Dirac masses via their Yukawa interactions with the Higgs field, except for three: the neutrinos. Nonzero neutrino masses require new degrees of freedom and, perhaps, new organizing principles.  

We currently know very little about these new degrees of freedom and discuss some examples in the next subsection. They could be very light or very heavy, bosons or fermions, charged or neutral. The new degress of freedom responsible for nonzero neutrino masses may also play a role in some of the other outstanding questions in fundamental particle physics and cosmology today, including grand unification, the matter--antimatter asymmetry of the universe, and the dark matter puzzle. 

Data are required in order to make qualitative progress. Neutrino theory and phenomeology are required in order to identify the different possible origins for neutrino mass and how/whether these can be distinguished and falsified.  

\subsection{What is the Mechanism Behind Non-Zero Neutrino Masses?}

Assuming the SM is an effective theory, it is productive to estimate the impact of hypothetical new ``heavy'' physics by exploring the consequences of nonrenormalizable interactions. There is only one dimension-five operator consistent with SM gauge invariance \cite{Weinberg:1979sa}:
\begin{equation}
    {\cal L}_5=-\frac{(LH)(LH)}{2\Lambda} + H.c.,
    \label{eq:L5}
\end{equation}
where flavor indices have been omitted. After electroweak symmetry breaking, ${\cal L}_5\to -m_{\nu}\nu\nu/2$, where $m_{\nu}=v^2/\Lambda$ are Majorana masses for the SM neutrinos. $v$ is the vev of the neutral component of $H$. Current information on neutrino masses, summarized in Section 5.4, translates into an effective scale associated to the dimension-five operator $\Lambda\sim 10^{15}$~GeV.

If the physics responsible for neutrino masses manifests itself at low energies via Eq.~(\ref{eq:L5}), we can safely state there are new degrees of freedom with masses at or below $10^{15}$~GeV. Eq.~(\ref{eq:L5}) also implies that lepton number (and its anomaly-free cousin, $U(1)_{B-L}$, baryon-number minus lepton-number) is broken once the new degrees of freedom are accounted for. 

There are three tree-level realizations of Eq.~(\ref{eq:L5}), assuming the existence of either new fermion fields or new scalar fields charged under the SM~WP\cite{Almumin:2022rml}. These are referred to as the Type I -- neutral fermion -- Type II -- triplet scalar -- and Type III -- triplet fermion -- versions of the seesaw mechanism. The presence of new charged particles implies that the new particles in the Type II and III seesaws are heavier than the weak scale, given their absence in LHC data. In the Type I seesaw, instead, the new ``heavy'' particles, as far as experimental constraints are concerned, can be as light as a few eV, as long as they are very weakly coupled to the SM. The eV lower bound is derived from oscillation experiments. 

\begin{figure}[htbp]
\begin{center}
\includegraphics[width=1\linewidth]{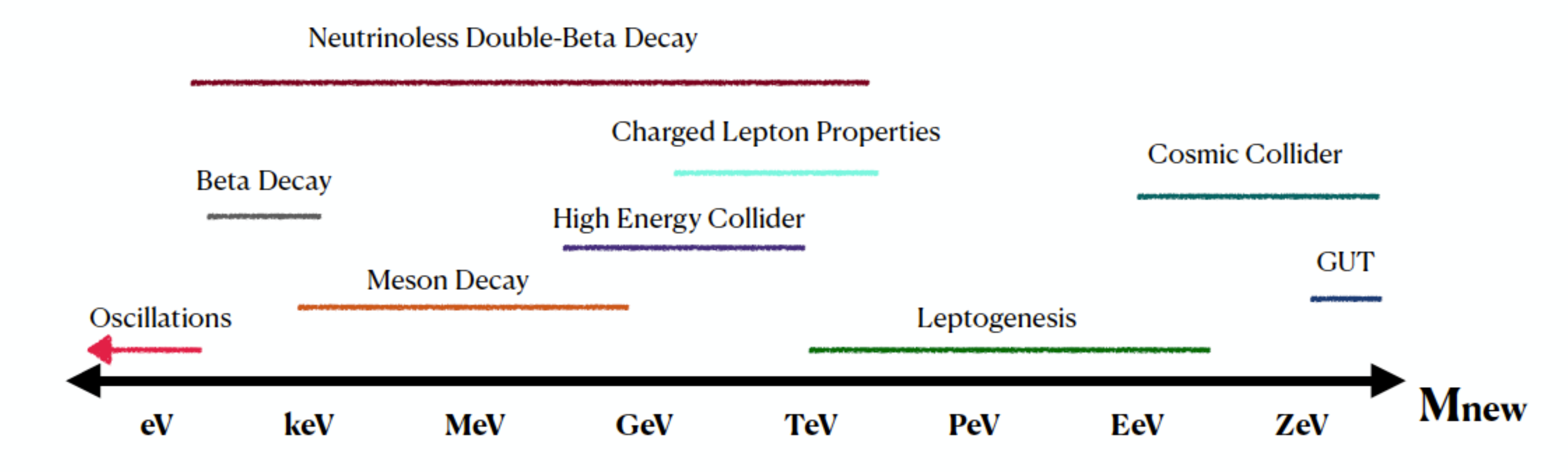}
\caption{The new physics responsible for nonzero neutrino masses can be anywhere between the eV and ZeV scale. Different scenarios lead to potential signatures in distinct particle and nuclear physics probes.}
\label{fig:scales}
\end{center}
\end{figure}

Instead, Eq.~(\ref{eq:L5}) might be generated once the new heavy particles are integrated out at the loop level~WP\cite{Almumin:2022rml}. In this case, even if the couplings that describe the interactions between SM and new particles are order one, the masses of the new particles are constrained to be much smaller than $10^{15}$~GeV in order to yield the known neutrino masses. Different new-physics scenarios leave imprints in different types of particle and nuclear physics probes, and may be related to other outstanding questions in fundamental particle physics. Some probes are illustrated in Figure~\ref{fig:scales}. A few concrete examples will be discussed in Section~\ref{sec:pheno}.

There is also the possibility that the dominant contribution to nonzero neutrino masses is not captured by Eq.~(\ref{eq:L5}). This is the case if there are SM gauge-singlet left-handed Weyl fermion fields $\nu^c$ (also known as right-handed neutrinos or left-handed antineutrinos) {\it and}  if lepton number (or, e.g., $U(1)_{B-L}$) is a fundamental symmetry of nature (or an excellent approximate symmetry). 

It is instructive to explore the general consequences of gauge-singlet fermions in more detail. At the renormalizable level, $\nu^c$ are only allowed to couple to SM degrees of freedom via Yukawa interactions,
\begin{equation}
{\cal L}_{4} = y_{\nu}LH\nu^c + H.c.,
\label{eq:L4}
\end{equation}
where again flavor indices have been omitted. After electroweak symmetry breaking, ${\cal L}_4\to m^D_{\nu}\nu\nu^c$, where $m^D_{\nu}=y_{\nu}v$ are the neutrino Dirac mass parameters. Gauge-singlet fermions are also allowed Majorana mass terms, i.e., one expects ${\cal L}\supset -M \nu^c\nu^c/2$, where $M$ is a new, unconstrained mass scale, a priori unrelated to the electroweak symmetry breaking scale. 

In the limit $M\gg m^D_{\nu}$, one can integrate out the $\nu^c$ fields and, at the light neutrino mass scale, one obtains Eq.~(\ref{eq:L5}). Indeed, this is the Type-I seesaw mechanism. Instead, in the limit $M\ll m^D_{\nu}$, neutrinos are pseudo-Dirac fermions -- quasi-degenerate, fifty-fifty mixtures of $\nu$ and $\nu^c$ -- with mass $m^D_{\nu}$. In this case, data translate into $y_{\nu}\sim 10^{-12}$. 

When $M$ vanishes, the global symmetry structure of the SM+$\nu^c$ Lagrangian  is enhanced: $U(1)_{B-L}$ is conserved.\footnote{This is also the case if $y_{\nu}$ were to vanish. In this case, however, neutrinos would be massless and the $\nu^c$ completely decoupled.}  In this case, neutrinos are Dirac fermions. This is qualitative different from the SM, where neutrinos are massless. In the SM, $U(1)_{B-L}$ is an accidental global symmetry and it is generically anticipated that new ultraviolet degrees of freedom will ultimately reveal that $U(1)_{B-L}$ is only approximate. If the neutrinos are massive Dirac fermions, $U(1)_{B-L}$ must be nontrivially imposed in order to ``explain'' why the $\nu^c$ Majorana masses vanish. In this case,  $U(1)_{B-L}$ must also be preserved by the unknown degrees of freedom that lurk in the ultraviolet. 

The hypothesis that neutrinos are Dirac fermions carries with it a few interesting theoretical puzzles. One is that it has been argued that quantum gravity effects do not allow for exact global symmetries~WP\cite{Draper:2022pvk}. If data are consistent with Dirac neutrinos, these quantum gravity arguments are incorrect,  $U(1)_{B-L}$ is a gauge symmetry, or $U(1)_{B-L}$ is explicitly very weakly broken. In the latter case, neutrinos are pseudo-Dirac fermions. As far as the Lagrangian at low energies is concerned, this translates into $0\neq M\ll m^D_{\nu}$ or the addition of Eq.~(\ref{eq:L5}) with very large $\Lambda\gg 10^{15}$~GeV. This scenario can be probed in searches for new neutrino oscillation lengths that are many orders of magnitude longer than the ones observed.  

Another theoretical puzzle is related to the magnitude of $y_{\nu}$, required to be smaller than the electron Yukawa coupling by at least six orders of magnitude. The tiny $y_{\nu}$ might be an indication that there is more to Eq.~(\ref{eq:L4}) than meets the eye. One simple hypothesis is that $\nu^c$ is charged under some dark-sector symmetry, invisible to all SM fields. In this case, Eq.~(\ref{eq:L4}) is forbidden by the dark symmetry but could be generated, via a higher-dimensional operator, if the dark symmetry were spontaneously broken (i.e., ${\cal L}_{4}$ is the low energy manifestation of the effective Lagrangian proportional to $(LH)(\nu^c\phi)$ where $\nu^c\phi$ is a dark-symmetry singlet and $\phi$ is a scalar field with a nontrivial vev). This scenario takes advantage of the neutrino portal to the hypothetical dark sector and allows one to explore neutrino--dark matter connections. For a concrete example, see \cite{Cherry:2014xra} and references therein and thereof.

\subsection{Neutrinos and the Flavor Puzzle}

The fact that there are three copies of every known fundamental matter field -- identical SM quantum numbers but qualitatively different masses -- is often interpreted as evidence that there are still-to-be-uncovered fundamental organizing principles and symmetries in nature.  Flavor symmetries, sometimes referred to as horizontal symmetries, appear to be specially promising. In a nutshell, the idea is to impose a new symmetry under which fermions from different generations are charged differently ~WP\cite{Almumin:2022rml,Gehrlein:2022nss,Altmannshofer:2022aml}. The hierarchy of the fermion masses and the patterns observed in the mixing matrices would be a consequence of the charges of the different fermion fields and the way in which the new horizontal symmetry is broken.

In the case of charged-leptons and neutrinos, the patterns observed in the leptonic mixing matrix are often a more natural fit for discrete flavor symmetries, which circumvent some of the challenges faced by continuous symmetries. This approach to flavor also has its limitations including a certain level of arbitrariness when it comes to choosing how the symmetries are broken. A recent, very promising direction that has attracted a lot of attention revolves around modular flavor symmetries. These do not suffer from the so-called vev-alignment problem, for example. We refer to~WP\cite{Almumin:2022rml,Gehrlein:2022nss} for a few more details.  

Flavor symmetries are often side-effects of physics in the far ultraviolet. Hence, the patterns observed in the neutrino masses and the leptonic mixing matrix may contain hints regarding, for example, the grand unification of fundamental fields and interactions or the string theory realization of fundamental physics. On the low-energy side, flavor symmetries manifest themselves in the form of relations among different Lagrangian parameters. Hence, precision measurements of flavor parameters constitute the most direct way of identifying whether there are hidden flavor symmetries and distinguishing one theoretical model from another. 

\begin{figure}[htbp]
\begin{center}
\includegraphics[width=0.4\linewidth]{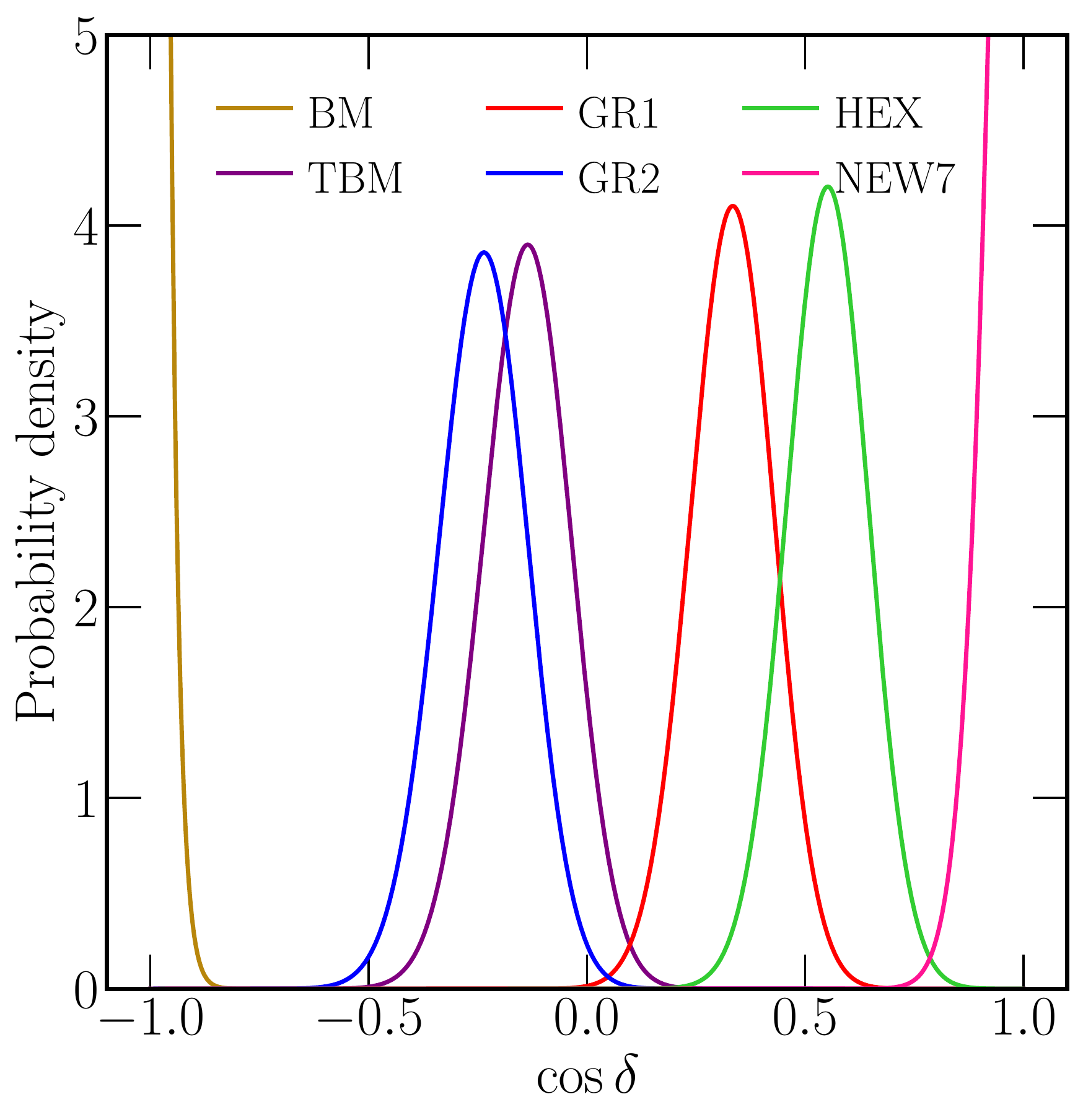}
\caption{Probability density as a function of $\cos\delta$  for various flavor models (different colored curves). From \cite{Everett:2019idp}, to which we refer for all the details.}
\label{fig:delta}
\end{center}
\end{figure}

Flavor symmetries also provide concrete precision-targets for next-generation neutrino oscillation experiments. One example is the currently unknown CP-odd phase $\delta$ that parameterizes CP-violating effects in neutrino oscillations. Different flavor models make qualitatively different predictions for $\delta$. The prediction can often be cast in the form of a theoretical probability distribution as depicted in Figure~\ref{fig:delta}, for various models, discussed in detail in \cite{Everett:2019idp}. The figure not only illustrates that different models make different predictions for $\cos\delta$ but also hints at how precise should $\cos\delta$ be measured if one were to successfully distinguish different scenarios.

Consequences of flavor symmetries also translate into relations among the different masses and mixing parameters (e.g., $\cos2\theta_{23}=\sin^2\theta_{13}$, see \cite{ParticleDataGroup:2020ssz} for the definition of the neutrino mixing parameters), and may involve, if grand unification is part of the model, quark and lepton mixing parameters (e.g., $\sin^2\theta_{13}=\sin^2\theta_c/\sqrt{2}$, where $\theta_c$ is the Cabibbo angle). These can be tested as precisely as the least constrained mixing parameter. For example, given our current understanding of neutrino mixing, the sum rule $\cos2\theta_{23}-\sin^2\theta_{13}=0$ is satisfied. At the three sigma level, $\cos2\theta_{23}\in[-0.24,0.19]$ while $\sin^2\theta_{13}\in [0.020,0.024]$. Hence, our ability to test this sum rule depends, virtually exclusively, on improving the precision with which we measure $\theta_{23}$. Other sum rules involve the CP-odd parameter $\delta$. For example, they might relate it to the equivalent phase $\delta_{\rm CKM}$ in the quark mixing matrix. $\delta_{\rm CKM}$ is known at the 6.5\% level ($\delta_{\rm CKM}=68.8^\circ\pm4.5^\circ$) hence, in order to test these relations, one would like to measure $\delta$ with equivalent precision.

\section{Neutrino Phenomenology}
\label{sec:pheno}

An avalanche of data on neutrino properties and interactions is expected in the next decade. The interpretation of these data, along with the development of new research questions, search strategies, and analyses, rely on the interplay of theoretical, phenomenological, and experimental efforts. Here we highlight several different probes that are guaranteed to speak to neutrino physics, identifying the interplay of theory and experiment. All of these are discussed, in more detail, in several Snowmass Topical Group Reports, including EF08, EF09, NF02, NF03, and RF5. While some overlap is unavoidable (and, in some cases, very useful), we intend the brief discussions here to complement those found elsewhere.  

\subsection{Neutrino Oscillations}

Neutrino oscillation experiments provide all affirmative information we have on neutrino masses. More, better neutrino oscillation data are expected in this and the next decade. Theory played a central role in the development of the formalism of neutrino oscillations, including the solution to the twentieth-century solar and atmospheric neutrino puzzles. This role is expected to persist and evolve as the oscillation probes grow more sophisticated and diverse. Currently, for example, the combination of different neutrino oscillation data is performed by different neutrino-phenomenology research groups. This theoretical research provides the best-fit values for the different oscillation parameters that are used by the particle physics community as a whole. Since our current picture of neutrino oscillations is the consequence of input from a large set of experiments, the relevance of this type of work is expected to remain very high. On the formalism side, questions include the limitations of the very simple yet incredibly successful oscillation picture -- which assumes that virtually all neutrino sources emit perfectly coherent superpositions of neutrinos with well-defined mass or perfectly incoherent mixtures of these particle -- and whether these can be probed, and how oscillations can be used to test some of the fundamental building blocks of basic Physics \cite{Takeuchi}. 

New neutrino oscillation data can also reveal more new physics in the neutrino sector. The role of theory here is also manifold. Phenomenological scenarios are required in order to define and compare the reach of new experimental proposals to new phenomena. These include, among others, new weaker-than-weak neutrino--matter interactions that modify neutrino production, detection, and flavor-evolution, new light neutral fermions that lead to the presence of new oscillation frequencies or to the apparent non-unitarity of the leptonic mixing matrix, or allowing for the possibility that neutrinos have a finite lifetime~WP\cite{Arguelles:2022xxa}. Theoretical and phenomenological work is required in order to compute these hypothetical new-physics effects in oscillation experiments and in order to transform these phenomenological scenarios into well-defined theoretical frameworks. The latter is required to compare the reach of neutrino experiments to other fundamental physics probes and to explore whether these models can address any of the outstanding questions in particle physics, including the dynamical origin of nonzero neutrino masses or the dark matter puzzle. There are, for example, different models that ``explain'' nonzero neutrino masses by adding, to the SM particle content, gauge-singlet fermions. Some of these were discussed in more detail earlier. If these new fermions are light enough, they can leave an imprint in oscillation experiments. Furthermore, the new-physics parameters are not generic but can be predicted assuming the new physics is directly responsible for nonzero neutrino masses. This allows one to set targets and translate, for example, failed searches for new neutrino states into meaningful information on the origin of non-zero neutrino masses. 

An interesting problem that is expected to become much more relevant in the next-generation of oscillation experiments\footnote{These are not only more precise but also include new oscillation channels, including tau-neutrino appearance~WP\cite{Abraham:2022jse}.} is to define how to perform model-independent searches for new phenomena using neutrino oscillations, and how to report the results of these searches. This will require a coherent neutrino theory and experiment research effort that targets the unique properties of oscillation probes and how these translate into surprises. Model-independent probes include experimentally over-constraining the neutrino oscillation parameters by performing qualitatively different experiments sensitive to -- assuming the three-massive-neutrinos paradigm is correct -- the same oscillation parameters and the construction of ``unitarity triangles,'' that are very familiar when exploring new phenomena in quark-flavor physics.   

In the future, new facilities aimed at studying neutrino oscillations, including neutrino factories, will pose more theoretical challenges and provide more opportunities for discovery~WP\cite{Bogacz:2022xsj}.

\subsection{The Short-Baseline Anomalies}

Neutrino theory also aims at interpreting unexpected results in neutrino experiments and computing the consequences of different interpretations. Currently, there is a handful of experimental data which, if taken at face value, cannot be explained within the three-massive-neutrinos paradigm. These are referred to, collectively, the short-baseline anomalies ~WP\cite{Acero:2022wqg}. 

There is no outstanding solution to the short-baseline anomalies~WP\cite{Acero:2022wqg}. It is possible that the correct answer is still to be uncovered, hopefully with the help of more and better neutrino data. Finding a definitive solution to the short-baseline anomalies is a high priority for neutrino physics this decade. There are, however, several different candidate-solutions that solve different subsets of the short-baseline anomalies, including several that postulate the existence of new physics. These include, for example, new neutrino states that participate in new interactions. These are not only important and interesting in their own right -- indeed, one of these models might be correct! -- and some can be related to, for example, the dark matter puzzle, but they also provide the ingredients necessary in order to define how to test, experimentally, the short-baseline anomalies. Recent results from MicroBooNE, and ongoing experimental analyses, rely on testing different, concrete hypotheses, including a new source of $\nu_{\mu}\to\nu_e$ flavor transitions \cite{MicroBooNE:2021ktl} or a very large rate for $\Delta \to N\gamma$ in nuclear environments \cite{MicroBooNE:2021zai}. New data from the short-baseline neutrino program at Fermilab, designed to, among other goals, resolve the short-baseline anomalies, will rely on more new-physics (and old-physics) hypotheses in order to reach their goals.

\subsection{Neutrino Scattering}

Since neutrinos interact only via the weak interactions, neutrino scattering experiments are especially sensitive to new, weaker-than-weak interactions mediated by hypothetical light or heavy new particles. Neutrino scattering experiments can reveal the existence of new interactions, new neutrino properties, or new particles. Neutrino scattering experiments, for the foreseeable future, are expected to be of the fixed-target type. The neutrino energies available for such experiments span more than 20 orders of magnitude. Here we concentrate on neutrino scattering and new phenomena. The complex issue of computing the details of neutrino--matter scattering in the SM is discussed in much more detail in Sec.~\ref{sec:xsec}.

At the lowest energies, there is research in detector technology capable of detecting the cosmic neutrino background via $\nu_e+X\to e^-+X'$, where $X$ is a beta-decaying nucleus. Such a measure would not only open a new window to the very early universe,\footnote{The cosmic neutrino background decoupled at temperatures around an MeV or 10 billion kelvin. Compare it to the cosmic background photons, which decoupled at 3000~K.} but measurements of the cosmic neutrino background are also expected to inform the nature of neutrinos -- Majorana fermions or Dirac fermions -- and are sensitive to the existence of new interactions or new light neutrino states \cite{PTOLEMY:2019hkd}. 

At higher energies, above tens of keV, below 10~MeV, there are solar neutrinos. These provide some of the most stringent bounds on the neutrino magnetic moment and are sensitive to, for example, new weaker-than-weak neutrino interactions.  

In the laboratory, coherent elastic neutrino--nucleus  scattering (CE$\nu$NS) was detected, using neutrinos from pion decay-at-rest, for the first time in 2017 \cite{COHERENT:2017ipa}. There are several collaborations around the world aiming at measuring CE$\nu$NS using both pion decay-at-rest and nuclear reactors as sources and different target nuclei. The CE$\nu$NS cross section, within the SM, can be computed with high precision and hence CE$\nu$NS measurements are sensitive to new interactions -- including those to which oscillation experiments are blind -- and allow one to investigate the properties of the nuclear targets and serve as low-energy electroweak precision observables~WP\cite{Abdullah:2022zue,Akimov:2022oyb}. At a relatively similar energy regime, access to intense neutrino sources at both nuclear reactors~ WP\cite{CONNIE:2022hna} and accelerators, allow one to perform high-statistics measurements of neutrino--electron scattering. This is another process that can be computed, in the SM, with high precision and hence serves as a low-energy electroweak precision observable and is especially sensitive to neutrino electromagnetic properties (for example, dipole moments). 

As an aside, massive neutrinos are guaranteed to have nonzero dipole moments. Assuming no new interactions other than those in the SM, the neutrino magnetic dipole moment is predicted to be around $10^{-20}~\mu_B$ ($\mu_B=e/2m_e$ is the Bohr magneton. $-e,m_e$ are the electron charge and mass). New interactions can lead to much larger magnetic dipole moments. However, since the electromagnetic dipole interactions are chirality violating, new physics that contributes to the neutrino magnetic dipole moment also contribute to the neutrino mass. The relationship between dipole moments and masses are model dependent and are different for Majorana and Dirac neutrinos. Ultimately, theory and model-building efforts are required in order to interpret and understand the reach and effectiveness of searches for nonzero neutrino electric and magnetic dipole moments.

Intense neutrino beams are produced with the help of particle accelerators. Current neutrino oscillation experiments make use of intense neutrino beams from pion and kaon decay-in-flight. These have energies that range between 100~MeV and several GeV. Neutrino scattering measurements at near-detector facilities are not only necessary for neutrino oscillation experiments but also allow a plethora of new-physics searches. These will be briefly reviewed in Section~\ref{sec:not-nu}.

At still higher energies, experiments at colliders produce, in a variety of ways, a large sample of neutrinos of all flavors ($\nu_e,\nu_{\mu},\nu_{\tau}$ and their antiparticles). These neutrinos can be measured in dedicated detectors located hundreds of meters away from the collision point, in the ``forward'' direction. The first such measurement was reported in 2021 \cite{FASER:2021mtu}. The study of high energy neutrino--matter scattering allows one to investigate the existence of new particles, new neutrino interactions, and new neutrino properties. These dedicated forward detectors, including their physics reach and capabilities, are under intense investigation~ WP\cite{Anchordoqui:2021ghd,Feng:2022inv,Bai:2022jcs}.  

The highest energy neutrinos come from the cosmos and are captured in dedicated gigantic neutrino detectors~ WP\cite{Ackermann:2022rqc}. IceCube has detected neutrinos with energies that approach $10^{7}$~GeV. In fact, IceCube has reported evidence for the so-called Glashow resonance \cite{IceCube:2021rpz}, $\bar{\nu}_e+e^-\to W^-$, and is sensitive to different, unique new-physics contributions to neutrino matter scattering. The scattering of those neutrinos with the atmosphere, which can be seen in cosmic ray detector arrays, allows access to unique information on deep-inelastic scattering.

\subsection{High Energy Colliders}

The rich environment found at high energy colliders, including the LHC, future very high energy hadron colliders, and future high energy electron and muon colliders allow unique information on the physics behind nonzero neutrino masses and access to neutrino properties and new neutrino interactions~ WP\cite{Han:2022qgg,MuonCollider:2022xlm,Bernardi:2022hny,ILCInternationalDevelopmentTeam:2022izu,Barzi:2022rax,Mekala:2022cmm}. 

As discussed earlier, neutrino masses imply the existence of new particles. If these are light enough and strongly-coupled enough, they can be produced and detected at high energy colliders. An important challenge for theory and phenomenology are to identify different new physics models that can be probed at colliders, what are their characteristic signatures, and how present and future collider data will inform different research directions. 

As a concrete example, right-handed neutrinos with Majorana masses of order tens to thousands of GeV can be produced at high energy colliders, just like ordinary neutrinos (via, for example, $W^*\to \ell \nu^c$ where $\ell$ are charged-leptons). These, in turn, decay via weak interactions: $\nu^c\to\nu Z$ or $\nu^c\to\ell W$, where the gauge bosons may be off-shell. Since $\nu^c$ are Majorana fermions, their decays do not respect lepton-number conservation and, for example, collider searches at the LHC often concentrate on lepton-number violating final states like $\ell^-\ell^-+X$ and no missing energy, where $X$ are visible states with zero lepton number (for example, jets). in the simple Type-I seesaw, the search for these neutrinos is tempered by the fact that the expected right-handed-neutrino production rates are tiny. Roughly speaking, one expects the production of the heavy neutrinos to be suppressed relative to that of light neutrinos by a factor $m_{\nu}/M$, around $10^{-12}$ for heavy neutrinos with masses of order 100~GeV. Different models allow for more strongly coupled new states, but these models bring about qualitatively different phenomenology. 

Lepton colliders allow access to different types of new neutrino physics and different channels. For example, heavy neutrinos with masses around tens of GeV are strongly constrained by LEP via failed searches for $e^+e^-\to \nu\nu^c$ followed by $\nu^c\to\ell+W^*$~\cite{DELPHI:1996qcc}. Lepton colliders also allow access to new lepton-number violating channels including initial states with nonzero lepton number (e.g., $e^-e^-$ or $e\gamma$ colliders). Muon colliders are especially intriguing for neutrino physics because of possible access to neutrinos in the initial state, either as real particles or as ``constituents'' of the muon. These allow, for example, one to consider measuring $\nu\bar{\nu}\to Z\to \ell^+\ell^-$ along with several other SM interactions, along with unique searches for new phenomena.

\subsection{Charged-Lepton and Meson Processes}

Charged-lepton and meson processes also provide unique information when it comes to the new physics that lies in the neutrino sector. The fact that these can be manipulated and produced in large quantities allows precision measurements of their properties and interactions.

Meson decays are sensitive to the presence of new, invisible light particles and hence can be used to constrain new neutrino states with masses below hundreds of MeV. For example, the decay $\pi\to\mu\nu^c$ for $\nu^c$ masses around tens of MeV leads to a final-state muon energy that is distinct from what is expected from the standard $\pi_{e2}$ decays and provides the most stringent constraint on the hypothesis that these particles, in this mass range, exists. New very light scalars that couple to neutrinos can also be constrained via, for example, $\pi\to\ell\nu\phi_{\rm new}$, where $\phi_{\rm new}$ is a new hypothetical state.  

Neutrinos and charged-leptons are intimately connected. The fact that neutrinos mix, for example, implies that lepton-flavor numbers are not conserved in nature so flavor-violating processes involving charged-leptons are also guaranteed to occur. In the absence of new interactions, however, the expected rates for charged-lepton flavor violating (CLFV) process, including $\mu\to e \gamma$, $\mu\to eee$, and $\mu\to e$-conversion in nuclei, are tiny (these processes are GIM suppressed and the neutrino masses are very small). New physics, including the interactions and degrees of freedom that may be responsible for nonzero neutrino masses, can lead to much larger rates for CLFV. Some models, in fact, are best constrained by CLFV searches. 

Theory and phenomenology are required in order to establish connections between neutrino masses and CLFV, interpret the results of different searches, identify under what circumstances they complement one another, and set targets and goals for future searches for CLFV.  An important feature of CLFV observables is their direct connection to flavor physics. Observations of CLFV processes would serve as vital information when it comes to testing different flavor models or underlying mechanisms behind the patterns observed in lepton masses and mixing. 

\section{Not-Neutrino Phenomenology for Neutrino Experiments}
\label{sec:not-nu}

Neutrino oscillation experiments consist of an intense neutrino beam with well defined flavor, a near-detector complex aimed at characterizing the neutrino beam and the interactions between neutrinos and the detector, and a very large detector located a long, strategic distance away from the neutrino source. These experiments aim at, for the most part, studying the phenomenon of neutrino oscillations, measuring neutrino properties, and looking for new physics associated to neutrinos. These experimental setups, however, allow other particle physics measurements and searches for new phenomena, many of which are only peripherally related to neutrinos. We discuss a few of them here, concentrating on the role of theory and phenomenology.

\subsection{Near-Detector Opportunities}

Near detectors at neutrino oscillation experiments are exposed to a huge flux of neutrinos. They consist of different detector technologies, sometimes mimicking the detector technology present in the far-detector site, sometimes taking advantage of the large neutrino fluxes in order to accomplish more specific or refined tasks. On the other hand, the pion decay-in-flight neutrino beam is a tertiary beam, as follows. A very intense proton beam impinges on a fixed target, producing a large number of mesons (depending on the proton energy, mostly pions and kaons, fewer charm mesons, etc). The charged mesons are guided and focused (or defocused) by magnetic horns before they decay into neutrinos and charged-leptons. These neutrinos define the neutrino beam. 

The fact that a large number of mesons is produced allows one to search for light, very weakly particles that might be emitted, with a small probability, when mesons decay. These, in turn, can find their way to the near-detector complex and either interact with the detector medium or decay in flight into visible particles. In sum, neutrino oscillation experiments are an excellent place to search for light, hidden sector particles. For example, if there is a dark photon $\gamma_{\rm dark}$, it may be produced in neutral pion decay $\pi^0\to\gamma\gamma_{\rm dark}$. Dark photons that make it to the near detector might decay into an electron--positron pair which can be identified and measured. 

There are several new-physics scenarios one can test at near-detector complexes. Many have received a lot of attention in the last few years and have been explored by the theory and phenomenology community~ WP\cite{Abdullahi:2022jlv,Berger:2022cab,Batell:2022xau}. These have also drawn the attention of the experimental community. There are many opportunities for future work, and the consequences of successful research programs can be especially impactful.  Theory is needed in order to identify the most promising new physics scenarios -- one needs to investigate connections to other outstanding puzzles or anomalies in particle physics -- and phenomenology is needed in order to identify the sensitivity of different facilities to the new physics and, perhaps more important, to identify whether there are ways to enhance the physics portfolio and capabilities of existing facilities.

\subsection{Far-Detector Opportunities}

Far detectors are large and located underground in order to minimize and control cosmic-ray induced backgrounds. These conditions are ideal for monitoring a large number of nucleons in search for proton decay and corresponding detectors and other baryon-number violating processes~WP\cite{Dev:2022jbf}. Such searches are the only definitive window available to pursue evidence for grand unification at very small distances. This fact was appreciated a long time ago and was the original motivation for the Kamiokande experiment. New detector technologies allow one to search for new decay modes and invite new search strategies. 

Large underground experiments are also excellent laboratories to study neutrinos from the Sun and the atmosphere. Theoretical efforts have also uncovered the fact that these detectors can be used to test some dark matter scenarios and to search for other exotic relics. As far-detector technologies evolve, theoretical and phenomenological work will be required in order to identify new capabilities and new search strategies. 

\section{Neutrinos in Astrophysics and Cosmology}
The past decade has seen important advances in neutrino astrophysics and cosmology. These advances inform us about the nature of the sources from which the neutrinos originate, but also inform us on the properties of neutrinos. Coordination with terrestrial experiments provides a unique method to study the fundamental interactions of neutrinos. In this section we highlight recent advances in high-energy and low-energy neutrino astrophysics, and anticipated results for the next decade. 

\subsection{High-energy neutrinos} 

\begin{figure}[htbp]
\begin{center}
\includegraphics[width=1\linewidth]{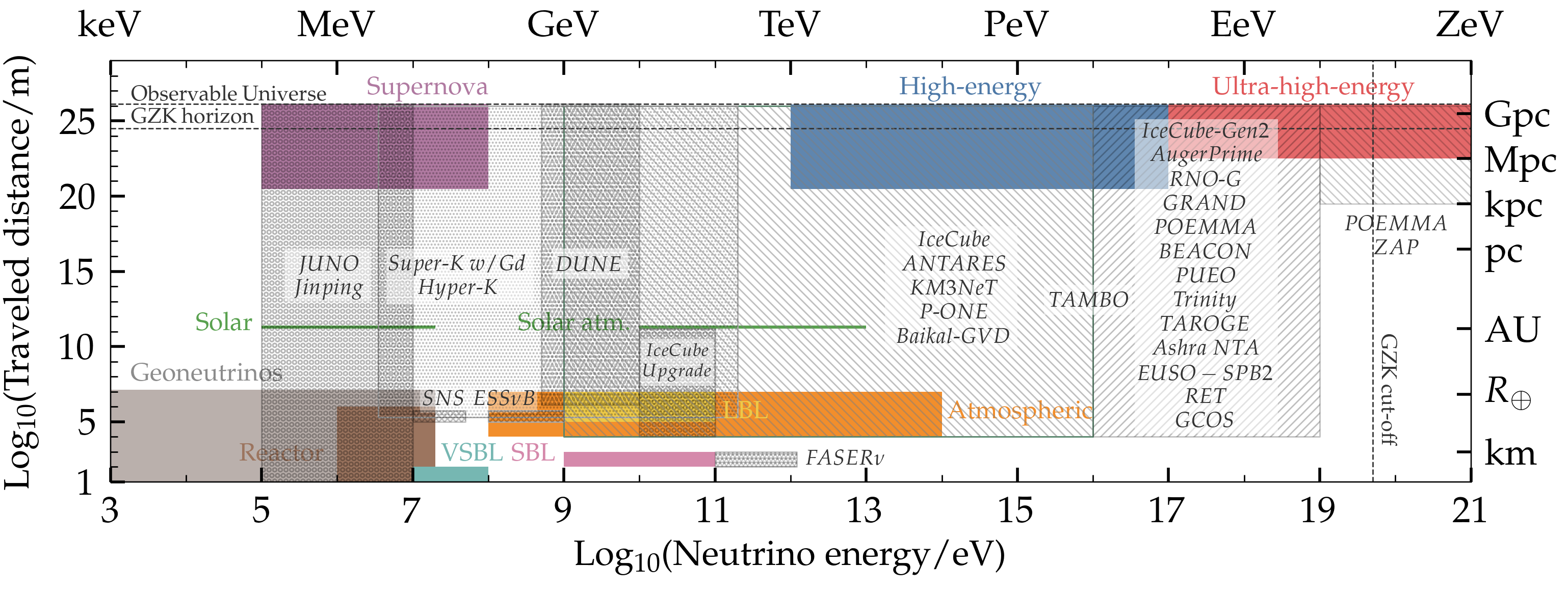}
\caption{Distribution of neutrino sources in energy and distance traveled to the detector, and present and future
experiments aimed at detecting them. We focus on high-energy and ultra-high energy neutrinos. From Ref.~WP\cite{Ackermann:2022rqc}.}
\label{fig:HE}
\end{center}
\end{figure}

Broadly speaking, high-energy neutrinos are directly or indirectly correlated with cosmic rays. The way in which they are produced and detected depends on ``how high'' is the high energy. They range from atmospheric neutrinos to the so-called GZK neutrinos. Figure~\ref{fig:HE}, from ~WP\cite{Ackermann:2022rqc}, summarizes the different neutrino sources and the rich experimental program aimed studying high energy cosmic neutrinos, as a function of the neutrino energy. Theory and phenomenology play a key role in using these different sources and energy scales in order to inform our understanding of neutrino properties and interactions, the properties of astrophysical objects and explosive phenomena, and the medium the neutrinos traverse. %

\subsubsection{GeV to TeV}
The IceCube DeepCore detector provides very high statistics atmospheric data at  energies from a few GeV to a few TeV, where new physics effects can be enhanced relative to standard ones due to the high neutrino energies and relatively enhanced matter effects. The high statistics can be used to reduce some of the systematic uncertainties associated with atmospheric neutrinos. In addition, the higher energies open up the possibility of measuring a larger subset of oscillation channels, including $\nu_\mu\to \nu_\tau$. Observing a large number of tau neutrinos and achieving better precision for measurements involving them would open access to large amounts of information about the standard three flavor neutrino scenario and numerous tests of physics beyond the standard model, from direct tests of leptonic unitarity in the neutrino sector to new interactions of tau neutrinos to interactions with dark sectors or other new particles, as well as particle astrophysics~WP\cite{Abraham:2022jse}.
An additional array of 7 strings, with 3m separation between DOMs, will be added to DeepCore within the following year. Future detectors like Hyper-K, KM3NET and DUNE will also be sensitive to this neutrino flux.
Testing neutrino oscillations in atmospheric neutrinos is extremely important for breaking degeneracies between standard and beyond-the-Standard-Model (BSM) parameters, and making sure that the three-neutrino paradigm is robust.

When it comes to atmospheric neutrinos, theory and phenomenology efforts play an important role in computing the atmospheric neutrino flux given what is known about cosmic rays, properties of the Earth, and cosmic-ray--atmosphere interations.

\subsubsection{TeV to PeV}

In the last decade, IceCube observed neutrinos with energies well above a TeV coming from outside our galaxy \cite{IceCube:2013low}. These provide a unique avenue to probe BSM effects, given the very long distances traveled by the neutrinos, as well as their ultra-high energies. Existing studies include the observation of the Glashow resonance, the first measurement of the HE neutrino-nucleon cross section, searches for neutrinos from dark matter annihilation and decay, and searches for new neutrino self-interactions. With steadily increasing event statistics in this decade and beyond, more subtle effects can be tested~WP\cite{Arguelles:2022xxa,Abraham:2022jse,Ackermann:2022rqc}.

The discovery of high energy astrophysical neutrinos in the last decade opened a new window into the Universe. A large population of dim sources likely explains the bulk of the neutrino observations. The neutrino energy density is comparable to that of ultra-high energy cosmic rays and gamma rays, hinting at the possibility of a common origin. With the improved statistics, sensitivity, and sky coverage offered by upcoming experiments, we can expect to expand our view of the neutrino sky, including firmly establishing neutrino sources. First hints of sources have already been reported \cite{IceCube:2018cha} and much more is expected in the next few years. In parallel, the first direct observations of gravitational waves and the discovery of new sources kicked-off the very exciting and promising area of multi-messenger astrophysics. With the advent of more sensitive neutrino experiments and dedicated follow-up facilities, more multi-messenger events at higher statistical significance are expected. 

In order to fully exploit current and future observations of high energy astrophysical neutrinos, dedicated, broad theoretical and phenomenological efforts are needed. Astrophysics and astrophysical modeling is needed in order to translate data into the properties of these extreme astrophysical objects, while neutrino particle and nuclear physics, including flavor-related effects, is required in order to properly describe neutrino production and relate it to the other messengers, extract particle physics information on, for example, neutrino properties. 

\subsubsection{Beyond a PeV}

It is not known how the diffuse astrophysical neutrino flux extends to higher energies. Finding out whether it cuts off at or beyond 100~PeV is crucial for understanding the physics behind ultra-high-energy cosmic rays and classifying sources. Recent advancements in cosmic ray physics physics have pushed expectations for the flux of ultra-high-energy neutrinos towards the low end. Today, it is widely anticipated that in order to probe the entire gamut of reasonable models for the origins of ultra-high-energy neutrinis, next-generation experiments need to target flux sensitivities of order
$10^{-10}$~GeV/cm$^2$/sr/s~WP\cite{Ackermann:2022rqc}.

For energies above several PeV, the strategies to hunt for ultra-high energy neutrinos change~WP\cite{Ackermann:2022rqc}. There are several experiments aiming to detect extremely high energy tau neutrinos, since these are not efficiently absorbed by the Earth. 
 Tau neutrinos would offer access to extremely high energies, where neutrino interactions have not yet been tested. They would provide understanding of new parameter space for both standard and beyond standard model scenarios, both by extending the energy range and because of the unique features in tau neutrino propagation~WP\cite{Abraham:2022jse}.

\subsection{Low-energy neutrinos} 

The cosmos also serves as a source of low-energy neutrinos. Some, like those from the Sun, are well known and have already provide a wealth of information not only about the inner workings of the Sun but also, quite famously, the neutrinos themselves. The impact of theoretical physics for elucidating the Solar Neutrino puzzle is well known and has already been mentioned. 

The detection of low-energy astrophysical neutrinos is an experimental and theoretical challenge onto itself. Theory and phenomenology is required in order to understand how these neutrinos ``look'' inside different detectors, and the interpretation of fugure low-energy astrophysical data will rely on phenomenological studies to understand different models and identify robust observables.  
 
\subsubsection{Supernova neutrinos}

Supernovae provide a unique laboratory for neutrino physics and astrophysics. Apart from the Big Bang, supernovae are the only environment in the universe where neutrinos are in or near thermal equilibrium conditions. If there were no neutrino oscillations, neutrinos would be expected to emerge from the core of a supernova with nearly Fermi-Dirac spectra associated to temperatures, for the $\nu_e$, $\bar{\nu}_e$, and muon/tau flavors, of $\sim 3,5,8$~MeV, respectively. Oscillations ``mix'' these different spectra in ways that are, still, not entirely known. Nonetheless, rough  predictions were broadly realized by the detection of twenty neutrinos from 1987A, confirming that neutrinos and antineutrinos of all flavors from core-collapse supernova explosions carry away 99\% of the energy associated with the burst. 

The detection of neutrinos from the next individual supernova contains invaluable information on neutrino properties, both standard and exotic~WP\cite{Hyper-Kamiokande:2022smq,Caratelli:2022llt}. One well known example is the neutrino mass ordering, which is expected to leave a robust imprint in the spectrum of supernova neutrinos. Extracting information, however, will require understanding the evolution of neutrino flavor inside the supernova explosion. This is a complex, multi-scale many-body theoretical physics problem that is yet to be resolved. In the last couple of decades, our understanding of neutrino oscillations inside the supernova has evolved dramatically (for one review, see \cite{Mirizzi:2015eza}) but we are far from the correct picture that captures all of the relevant scales in density, energy, and time.   

More generally, the neutrinos emitted from supernovae provide an ideal test of dense matter conditions, which have unique impact on collective neutrino flavor transformations, and contain information on a variety of fundamental astrophysics questions, including the synthesis of heavy elements and the physics behind the supernova explosions themselves. 

Neutrinos from supernovae at cosmological distances can be studied through the diffuse supernova neutrino background (DSNB). The DSNB is the predicted background from core-collapse supernovae throughout the history of the universe. Though the DSNB has not been directly detected, there are strong upper bounds on the $\bar{\nu}_e$ component of the flux from Super-Kamiokande~\cite{Super-Kamiokande:2011lwo}. With theory input, both of the particle physics and astrophysics type, measurements of the DSNB can be translated into information on neutrino properties and other BSM physics.

\subsubsection{Solar Neutrinos}

For the past half-century, solar neutrinos have provided invaluable information on the properties of neutrinos and on the physics of the solar interior. The combination of all solar neutrino data with terrestrial experiments favor the LMA-MSW solution to neutrino flavor transformation from the Sun to the Earth. At low energies, $\lesssim 5$ MeV, vacuum oscillations describe the neutrino flavor transformation, and the electron neutrino survival probability is around 60\%. At energies $\gtrsim 5$ MeV, matter-induced transformations describe the flavor transformation, with a corresponding survival probability of around 30\%.

Even with the tremendous theoretical and experimental progress in the field of solar neutrinos, there are still some outstanding questions that surround some of the data. For example, three experiments (Super-Kamiokande, SNO, and Borexino) that are sensitive to electron recoils from neutrino-electron elastic scattering find that at electron recoil energies of a few MeV, the data are $\sim 2\sigma$ discrepant relative to the prediction of the best-fitting LMA-MSA solution. This may be indicative of new physics, perhaps in the form of sterile neutrinos. In addition, the recent measurement of the solar mass-squared difference from solar neutrino data, in particular from the day-night Super-Kamiokande data~\cite{Super-Kamiokande:2016yck}, is slightly discrepant relative to that measured by KamLAND~\cite{KamLAND:2010fvi}. This may also be explained by novel physics in the neutrino sector. New data from JUNO is expected to add non-trivially to this discussion. 

Another outstanding question relates to how the measured flux informs the physics of the solar interior. Modeling of solar absorption spectra and heliosiesmology data suggests a lower abundance of metals in the solar core, i.e. a low-Z standard solar model (SSM)~\cite{Asplund:2009fu}. This is in comparison to the previously-established high-Z SSM~\cite{Grevesse:1998bj}. Though some sets of solar neutrino data favor a high-Z SSM~\cite{Borexino:2017rsf}, a global analysis of all solar neutrino fluxes remains inconclusive~\cite{Bergstrom:2016cbh}.

An interesting more recent development is expectations that next-generation dark matter direct-detection experiments will be able to measure the flux of low-energy solar neutrinos, including $pp$-neutrinos~WP\cite{Aalbers:2022dzr}. These provide complementary information on neutrino properties and astrophysics.  

\subsection{Cosmological Neutrinos} 

The Big Bang produced a thermal spectrum (now very cold) of neutrinos and antineutrinos of all species. These contain a wealth of information on particle physics, astrophysics and cosmology. The information is encoded in different large-scale observables. Theory and phenomenology play a central role in extracting and interpreting this information.

\subsubsection{Early Universe: Cosmic Microwave Background  and Recombination} 

Neutrinos contribute a significant component of the radiation density in the universe, and so are a substantial component of the energy density in the early universe. Neutrinos contribute to the evolution of the universe through both their mean density and their fluctuations. Both of these effects can be observed through precision measurements of the cosmic microwave background (CMB) and probes of the recombination of the universe. The energy density of neutrinos is expressed in terms of the effective number of neutrino species, $N_{\rm eff}$, which is proportional to the energy density of neutrinos to that of photons~WP\cite{Abazajian:2022ofy}. Assuming that neutrinos decouple instantaneously prior to electron-positron annihilation, then in the SM, $N_{\rm eff} = 3$. A deviation of $N_{\rm eff}$ can be interpreted as new neutrino physics or new light degrees of freedom~WP\cite{Dvorkin:2022jyg}. 

From their impact on the the temperature and polarization power spectra, the Planck satellite has measured a value of  in a $N_{\rm eff} = 2.92_{-0.19}^{+0.18}$, consistent with the SM prediction. This constraint is expected to be improved by factors of a few with future CMB experiments~WP\cite{Chang:2022tzj}. 

\subsubsection{Late Universe: Clustering and large-Scale Structure} 

Neutrinos affect the growth of structure through a change in the background expansion rate of the universe, and through their significant thermal velocities. The overall effect of massive neutrinos is to suppress the growth of structure at small scales, corresponding to wavenumbers $\lesssim 1$~k [$h$/Mpc]. Neutrinos induce a scale-dependent effect on the matter power spectrum that is difficult to mimic with other physics. Upcoming experiments such as Euclid and the Vera Rubin Observatory will be sensitive to the sum of the neutrino masses through their imprint on scales smaller than the free-streaming scale~WP\cite{Abazajian:2022ofy}. Extracting neutrino masses from these cosmic surveys is a nontrivial exercise that requires dedicated theoretical analyses and computations. 

\subsection{Absolute Neutrino Mass Measurements}

Neutrino oscillation experiments place a lower bound on the sum of neutrino masses, $\sum m_{\nu}\gtrsim 0.06$~eV~\cite{Gonzalez-Garcia:2021dve}. On the other hand, the inferred  upper limits on the sum of the neutrino masses from CMB measurements and those of other large-scale properties of the universe are are strict as $\sum m_{\nu}<0.12$~eV~\cite{Planck:2018vyg}. At face value, it seems that a nonzero neutrino mass measurement from cosmic surveys is a matter of time~WP\cite{Abazajian:2022ofy}. Concurrently, precision measurements of the $\beta$-spectrum of tritium decay are directly sensitive to kinematical consequences of nonzero neutrino masses. The Karlsruhe Tritium Neutrino (KATRIN) experiment currently binds the sum of the neutrino masses to be $\sum m_{\nu}<2.4$~eV~\cite{KATRIN:2022ayy}. In the future, up to an order of magnitude improvement can be expected \cite{Formaggio:2021nfz}. 

The interplay of these different probes of neutrino mass goes beyond simply measuring the neutrino masses. The real power, and the complementarity of the different types of  measurements, lies in investigating whether different observables are consistent with one another. Since the cosmological measurement of $\sum m_{\nu}$ is indirect, inconsistencies between cosmology, oscillations, and kinematical measurements will, mostly likely, be an indication that our understanding of the time-evolution of the universe, including its ingredients and their interactions, is incomplete. 

\section{Neutrino Cross Sections} 
\label{sec:xsec}

Neutrino physics is entering a precision era in which measurements of neutrino oscillations, 
astrophysical neutrinos from supernovae and other sources, and coherent neutrino scattering
will provide insight on the nature of neutrino masses, the presence of CP violation, and
more exotic new physics in the neutrino sector and beyond. Nuclei are used for these high-precision
tests of the SM and for searches of physics BSM. Without a thorough understanding of the 
way neutrinos interact with nuclei, we will not be able to meaningfully interpret the 
experimental data nor can we disentangle new physics signals from underlying nuclear effects.
Maximizing the discovery potential of increasingly precise neutrino experiments will 
require an improved theoretical understanding of neutrino-nucleus and neutrino-nucleon 
cross sections over a wide range of energies that uses a combination of lattice QCD (LQCD), 
nuclear many-body effective theories, phenomenological models, and neutrino event generators to 
make reliable theoretical predictions for experimentally relevant nuclei~WP\cite{Ruso:2022qes}.
This is a 
particularly challenging problem because of the wide range of energies and momenta 
involved in these experiments, from quasi-elastic (QE) scattering dominated by single-nucleon 
knockout processes, to the pion production region, and eventually to the 
deep inelastic scattering (DIS) region where the neutrino can resolve the individual quark constituents of the nucleon. 
Each of these regimes requires knowledge of both the nuclear ground state and the 
electroweak coupling and propagation of the struck nucleons, hadrons, or partons. 
The range of challenges is extreme; QE scattering and DIS  are conceptually the easiest to understand, but ultimately we would like to be able to predict both inclusive and exclusive cross sections across a wide range 
of kinematics. Theoretical efforts on neutrino scattering at the interface of high-energy and nuclear physics 
will be critical for achieving reliable cross-section predictions across the range of energies relevant to 
current and planned neutrino experimental programs.
%
\begin{figure}[htbp]
  \begin{center}
    \includegraphics[width=0.6\textwidth]{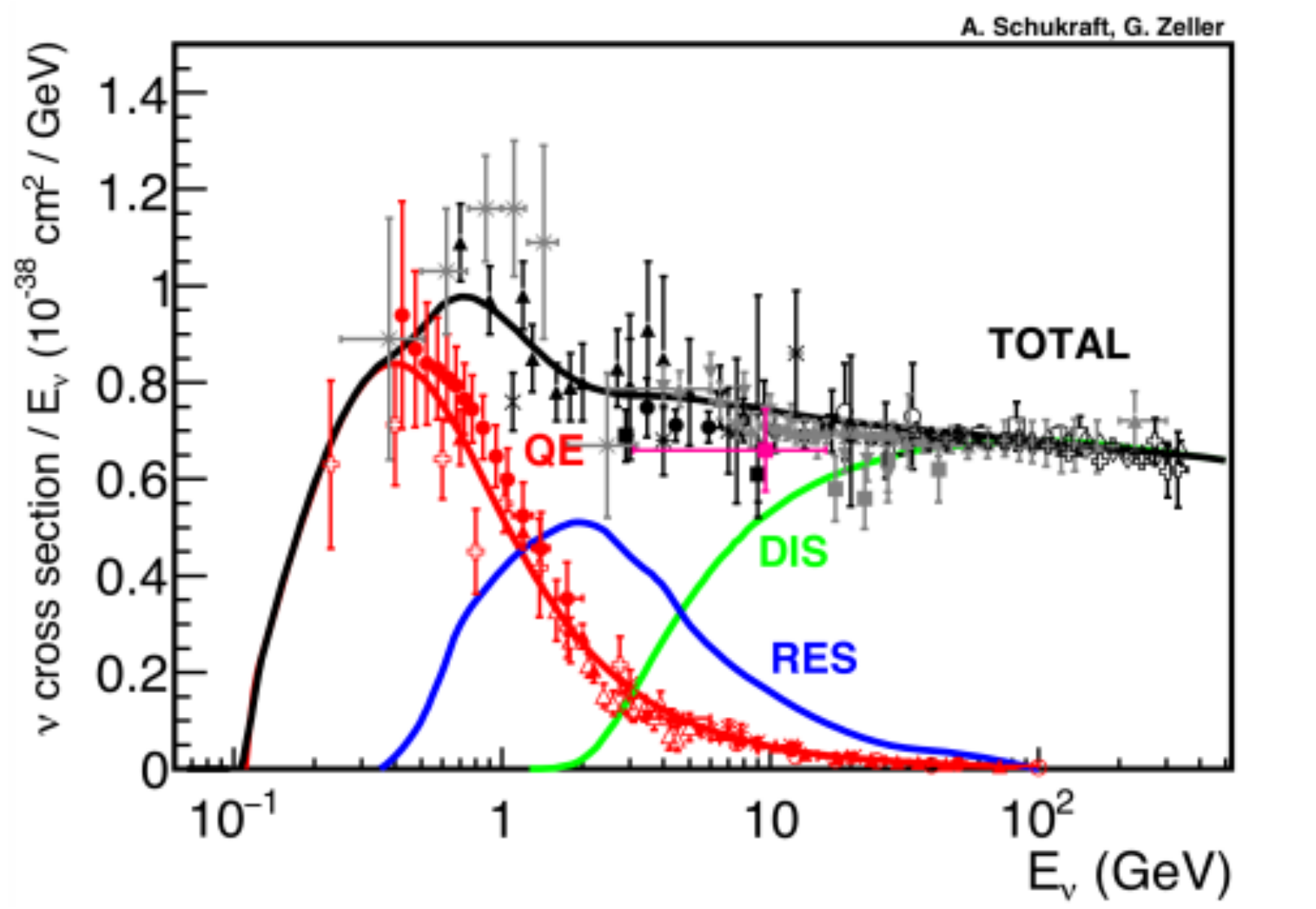}
 \end{center}
  \caption{\small Total neutrino per nucleon charged current cross section. Figure adapted from Ref.~\cite{Formaggio:2012cpf}.
\label{fig:DUNE:flux}}
\end{figure}

\subsection{The needs of the neutrino experimental program}

The planned neutrino experimental program probes a wide range of open physics questions. In many instances,
nuclear uncertainties constitute an obstacle to potential discoveries.
Broadly, there are three main energy regimes of interest. At low energies, there are precision 
measurements of coherent elastic neutrino scattering (CE$\nu$NS) and astrophysical sources 
(supernova neutrino bursts, SNB). At energies around 1 GeV, measurements of neutrino oscillation,
searches for exotic physics (sterile neutrinos, light dark matter), and searches for BSM 
processes, including proton decay, are made with atmospheric and accelerator based neutrino sources. 
At very high energies, there are astrophysical neutrino searches and precision tests of the SM
where detector response, resolution and statistical precision are the limiting factors more than
theoretical uncertainties. 

At low neutrino's energies ($E_\nu\sim1$--100~MeV), as CE$\nu$NS experiments continue to improve their experimental precision, 
more precise theory calculations of neutron distributions in the nuclear targets, which dominate 
the theoretical uncertainties, are required. In the same energy window,
calculations of MeV-scale exclusive scattering involving nuclear ground
and excited states will be needed to reconstruct the energies of astrophysical neutrinos from
SNB. 
In the intermediate energy regime ($E_\nu \sim 0.1$--20~GeV), especially important for the next-generation long-baseline oscillation projects DUNE and Hyper-Kamiokande, QE scattering, multi-nucleon dynamics, 
resonant processes (RES), DIS, and the transition region play an increasingly 
important role for future oscillation measurements (see Fig.~\ref{fig:DUNE:flux}.) 
Accelerator-based (anti)neutrino sources and atmospheric neutrinos have a broad energy spectrum.
Predictions for experimentally observable hadronic final states are required to reconstruct the incident
neutrino energies used in oscillation analyses, but several different reaction channels
contribute to particular final-state event rates. For each channel, well-grounded theoretical predictions are needed 
to assess event rates and uncertainties. 

The experimental neutrino program is in need of  accurate theoretical calculations of neutrino-nucleus and neutrino-nucleon
cross sections, both inclusive and exclusive,  with quantified theoretical errors
to ensure a robust implementation of interaction models in experiments.
Studies of electron scattering from nuclei and nucleons~WP\cite{Ankowski:2022thw} as well as photoproduction processes and pion-nucleus scattering are extremely useful
to validate the theoretical models and should also be pursued vigorously. New neutrino-hydrogen/deuterium measurements would greatly inform theoretical models and help 
validations of phenomenological models~WP\cite{Alvarez-Ruso:2022ctb}.
A critical need for the future will be close collaborations between theory and experimental groups.  
The expertise needed to carry out theoretical work in neutrino scattering comes from both high-energy 
and nuclear physics. 
Collaborations 
of theorists working on specific topics
across nuclear and high-energy physics will be extremely useful to enable progress in this area. 

\subsection{Coherent elastic neutrino-nucleus scattering}

\par Coherent elastic neutrino-nucleus scattering (CE$\nu$NS) is a process in which neutrinos scatter on a nucleus which acts as a single particle.  Though the total cross section is large by neutrino standards, CE$\nu$NS has long proven difficult to detect, since the deposited energy into the nucleus is $\sim$ keV. In 2017, the COHERENT collaboration announced the detection of CE$\nu$NS using a stopped-pion source with CsI detectors, followed up the detection of CE$\nu$NS using an Ar target. The detection of CE$\nu$NS has spawned a flurry of activities in high-energy physics, inspiring new constraints on BSM physics, and new experimental methods and dark matter direct detection. The CE$\nu$NS process has important implications for not only high-energy physics, but also astrophysics, nuclear physics, and beyond. Present experiments such as COHERENT are informing theory, and also how future experiments will provide a wealth of information across the aforementioned fields of physics. 

\par We are just at the very beginning of an exciting era in  CE$\nu$NS research. There is a multi-faceted experimental effort on-going around the world to expand upon the COHERENT measurements and to study \cevns\, using different neutrino sources and detector technology both as a means to study the CE$\nu$NS interaction itself and to probe other aspects of physics. Given the broad scientific applications of CE$\nu$NS, and its complementarity to many different aspects neutrino physics, it will be an important aspect of the neutrino physics program in the coming decade. 

\subsection{LQCD and inputs for neutrino scattering}

Many-body nuclear theories, where nucleons are taken as the relevant degrees of freedom,
rely on experimental data in the few-nucleon sector as well as LQCD and 
phenomenological calculations for single- and few-nucleon properties. For example, electron scattering experiments
provide crucial data on nucleonic elastic vector-current form factors, while nucleon elastic axial form
factors, that also enter neutrino scattering, are known less precisely from experiments and need to be supplied
by theoretical calculations. LQCD can provide first principles calculations of non-perturbative inputs to nuclear many-body effective theories 
({\it i.e.}, nucleonic elastic form factors, transition form factors, pion-production amplitudes, and two-nucleon responses) 
and nucleon parton distribution functions (PDFs), and sustained computational resources for achieving targeted precision goals for this wide range of quantities will be essential. Additionally, future LQCD calculations could serve as benchmarks
for phenomenological models of resonant scattering in the transition 
region between the low- and high-energy regimes. Several LQCD results for nucleon 
electroweak elastic form factors are reaching the percent precision over statistical and systematic uncertainties, providing already valuable inputs to 
nuclear many-body theories. Complementary to direct lattice calculations, there is also an 
extensive literature on determinations of the electromagnetic and axial form factors of 
the nucleon based on phenomenological fits and theoretical models. There are extremely valuable
synergies between the available lattice and phenomenological/model-based methods that can 
help extend or benchmark one approach off the other. In the low-energy regime, 
LQCD can play an important role in constraining two-nucleon dynamics ({\it i.e.}, two-nucleon correlations
and currents) by providing the low-energy constants entering the two-nucleon operators
used in nuclear effective theories as well as systematically controlled predictions for the more poorly known 
energy- and momentum-dependence of electroweak two-nucleon currents. See also WP\cite{Ruso:2022qes}.

Above the pion-production threshold, both resonant and non-resonant pion production processes 
must be theoretically understood at the ten-percent level in order to achieve few-percent overall 
cross-section uncertainties~\cite{DUNE:2021tad}. LQCD calculations of prime importance to be pursued in the coming years include 
the resonant $N \rightarrow \Delta$ and $N\rightarrow N^*$ transition form factors, and more generally, 
the $N\rightarrow N\pi$ pion-production amplitudes. These calculations will be extremely 
valuable to further validate phenomenological models currently employed to describe these processes, 
such as, the highly sophisticated dynamical coupled-channels model used to describe photo-, electron- and 
neutrino-pion production, and to constrain the neutrino induced pion production operators 
derived form effective field theories (EFTs)~WP\cite{Ruso:2022qes}.

In the high-energy DIS region, hadronic cross sections
factorize into partonic cross sections calculable with perturbative QCD and light-cone
structure functions such as PDFs that must be determined through global fits to 
experimental data and/or non-perturbative calculations.  Lattice QCD studies of PDFs
are rapidly maturing and can also provide insight on aspects of nucleon and nuclear structure
functions relevant to neutrino scattering~WP\cite{Constantinou:2022yye}. 
How to correctly describe the transition between the pion production 
and the region dominated by the DIS is an open question that will 
need to be carefully addressed in the future and is discussed in more detail below.

\subsection{Nuclear many-body theory approaches}

Microscopic nuclear approaches aim at describing the structure and dynamics of 
atomic nuclei as emerging form the interactions among nucleons. The latter are
dictated by QCD and its symmetries. At low energies, the nucleus is modeled 
as a collection of $A$ non-relativistic nucleons correlated in pairs and triplets. 
External electroweak probes (electrons, neutrinos, photons, \dots) 
interact with individual nucleons inside the nucleus via one-body 
current operators and  with pairs and clusters of correlated nucleons via two- and many-nucleon
electroweak currents. These many-body operators are constructed from 
phenomenological models and, as of recently, from chiral EFTs. EFTs are a low-energy approximation of QCD 
that use bound states of QCD 
(nucleons, pions, $\Delta$'s, $\dots$) as degrees of freedom. EFTs allow 
for a systematic low-momentum expansion of nuclear many-body interactions and currents.
Due to the low-momentum expansion parameter, it is possible in principle
to evaluate nuclear observables to any degree of desired accuracy
with an associated theoretical error coming from the truncation in the EFT's
expansion parameter. 
As discussed above, elementary amplitudes, including elastic and transition nucleonic form-factors,
as well as low energy constant entering the EFTs many-body interactions and currents, 
are the main inputs to the nuclear models and they  are
provided by the data (if available) or by theoretical LQCD and phenomenological  calculations.
The microscopic or {\it ab initio} approach yields 
a coherent picture of the nucleus and its properties (both static and dynamical), 
and indicates that many-nucleon effects in both nuclear interactions 
and electroweak currents are essential to accurately explain the data. 
To solve for the nuclear many-body structure and dynamics, 
many sophisticated computational schemes are being developed and, in recent
years, also due  to the increased computational resources, 
the  microscopic approach has been extended to the study of medium-mass nuclei.

In the low-energy regime, nuclear many-body theory plays an important role in 
supporting the experimental \cevns\,  program by providing accurate calculations
of nuclear weak form factors. This is also the energy regime of single- and 
double-beta decay (including neutrinoless double beta decay discussed below)
for which nuclear theory has to provide the corresponding nuclear matrix elements.
At these energies, neutrinos can also inelastically scatter from nuclei, exciting low-lying 
nuclear states and at a bit higher energies ejecting nucleons from the nucleus. These 
inelastic processes play a key role in setting the nuclear environment in core-collapse 
supernovae and neutron star mergers, for example. Measuring these processes
in terrestrial detectors enables one to obtain the flavor- and energy-dependent neutrino flux
from supernovae, which can inform us about the internal dynamics of the astrophysical
site. At higher excitation energies, also collective modes in the nucleus can be excited and
eventually the QE regime is reached. Modeling these inelastic processes from the
theoretical point of view is more challenging than calculating ground state properties and
often more challenging than calculations of inclusive neutrino cross sections.

In the QE region, the {\it ab initio} community has been, to date, primarily focused on 
calculations of inclusive processes induced by electrons and neutrinos 
scattering from nuclei.  These calculations, due to the high computational
cost, have been limited to light nuclei. One of the next goals is then 
to develop new algorithms that allow for cross-section calculations
of larger nuclear system without losing the resolution acquired in the {\it ab initio} 
framework, that is without losing the important many-body effects required to accurately explain the data. 
Other very important developments to be carried out in the near future are the 
inclusion of relativistic effects and of exclusive channels.
Microscopic nuclear models
that are based on the factorization of the final states are suitable for the inclusion of these effects.
In the future, it may be also possible to perform calculations of nuclear response functions
through quantum computers. 

While rigorous, the drawback of the {\it ab inito} approach is its high computational cost
which currently limits its applicability to light to medium-mass ($A\sim 40$) nuclei. 
There are several nuclear many-body models that have the advantage of being 
less computationally expensive and, in many cases, able to accommodate
for both exclusive channels and relativistic effects in a straightforward way. 
To this class belong, for example, semi-phenomenological factorization approaches
like mean-field approaches. These methods successfully describe
both inclusive and exclusive channels induced by electrons and neutrinos in a wide kinematical range.
Lastly, a priority of this program is to determine reliable theoretical uncertainties to the 
calculated neutrino-nucleus cross sections. These can be readily 
attainable within many-body EFT-based approaches supplemented by LQCD inputs for few-nucleon 
systems. When using more phenomenological approaches, model dependence  as well as
variations with respect to the adopted nuclear computational methods should be estimated. See also WP\cite{Ruso:2022qes}.

\subsection{Neutrino-induced shallow and deep inelastic scattering}

An important challenge for achieving precise neutrino-nucleus cross-section predictions will be reliably bridging
the transition regions between low- and high-energy theories, which use different degrees of 
freedom to describe neutrino-nucleus interactions. The transition region between the $\Delta(1232)$ 
resonance excitation and DIS is referred to as the shallow inelastic scattering (SIS) region~WP\cite{Ruso:2022qes}\cite{NuSTEC:2017hzk} and  
can contribute significantly to the determination of neutrino oscillation parameters. The
science of SIS is poorly understood both theoretically and experimentally and 
encompasses the transition from strong interactions described in terms of hadronic 
degrees of freedom to those among quarks and gluons described by perturbative QCD.

Neutrino-scattering simulations often describe this transition using PDFs empirically 
extrapolated from the DIS region to lower values of of the invariant mass, $W$, of the final hadronic system,
and four-momentum transferred squared, $-Q^2$. Duality arguments constrain the inclusive
cross section but do not predict the specific particle content of the final state. 
Therefore, efforts to extend the description in terms of quarks and gluons towards lower $W$ and $Q^2$ by including
higher-twist corrections should be complemented with a realistic modeling of the 
SIS region using hadronic degrees of freedom. Progress in this direction has been significant
but is hindered by the lack of experimental information about the axial current 
for inelastic processes at non-zero $Q^2$. For example, extrapolations of 
vector-current structure functions from the dynamical coupled-channel model
of the resonance region to the DIS region nicely approach the corresponding structure 
function results obtained from partonic descriptions valid at high energies, however analogous 
extrapolations of axial-current structure functions between resonance
and DIS regions do not agree. New neutrino-hydrogen/deuterium DIS measurements would greatly
inform theoretical models and help more precisely determine the combinations of parton
distribution functions (PDFs) relevant to neutrino scattering as well as benchmarks for
validating phenomenological models of the transition region~WP\cite{Alvarez-Ruso:2022ctb}. 
Additionally, the process of converting from partonic degrees of freedom back to specific hadronic final states
is referred as hadronization. Significant effort has been invested in developing models for hadronization for
collider experiments. However, these models are tuned to experimental data with $Q^2 > 10$ GeV${}^2$,
significantly higher than the typical values in neutrino experiments. Experimental data and theoretical inputs will
be required to develop hadronization models relevant for this energy region~WP\cite{Campbell:2022qmc}.

Modern experiments with (heavy) nuclear targets have provided and will keep providing valuable 
information on these issues, but the presence of nuclear effects 
such as Fermi motion, Pauli blocking, long- and short-range correlations, 
two- and three-body currents and, very significantly, 
final-state interactions tends to blur the information required to 
refine the hadronic description in the way outlined above.  
Since a large fraction of events in NOvA and DUNE, 
and in atmospheric neutrino measurements at IceCube-Upgrade, 
KM3NeT, Super- and Hyper-Kamiokande, are from the SIS and DIS regions, 
there is a definite need to improve our knowledge of this physics. 
Detailed phenomenological studies of SIS, 
DIS, and transition regions will be needed to obtain consistent
models of neutrino scattering.

\subsection{Implementations within neutrino event generators}

Across the broad range of energies of interest for current and future
investigations of neutrino physics, realistic simulations of neutrino
interactions are a critical ingredient for the design, execution, and
interpretation of experimental analyses. These simulations are generally
carried out using Monte Carlo techniques implemented within computer programs
known as \textit{event generators}. Advances in our collective theoretical
understanding of neutrino interactions will be essential to the progress 
of the field. However, unless these
advances are appropriately translated into improvements to neutrino event
generators, the benefit of theory efforts on experimental precision will be
severely limited at best~WP\cite{Campbell:2022qmc,Ruso:2022qes}.

The direct involvement of theorists within the neutrino event generator community is very limited, and development is mostly lead by experimentalists. This leads to significant delays in the adoption of new nuclear models into the code bases, creating an additional barrier between new nuclear models and data comparisons. In order to meet the precision needs of current- and next-generation neutrino experiments, involvement in event generators from theorists needs to increase. Additionally, the modeling of the propagation of outputs of the primary interaction through the nucleus, known as final state interactions, is solely the responsibility of the event generators. The modeling of these final state interactions is currently one of the dominant systematic uncertainties. Developing a rigorous theoretical description of this process and the connection to the primary interaction are essential to meeting the precision requirements of DUNE. Furthermore, to take advantage of the general purpose near detector of DUNE, event generators will need to develop methods to efficiently simulate BSM scenarios through the use of an automated pipeline, instead of the current approach of implementing each model by hand.

While the development of neutrino event generators is technically demanding,
sociological and organizational challenges are currently the greatest
hindrances to progress. Participation in generator-related activities is poorly
incentivized for both theorists and experimentalists, and opportunities to
pursue neutrino generator development as one's primary research activity are rare. 
Work on neutrino event generators is largely driven by idiosyncratic 
short-term needs of individual experiments and interests of small theory groups. 
A need for greater coordination, prioritization, and sustained support of such activities
as well as improved career incentives for physicists working on generators are
widely recognized in the neutrino scattering community.
A notable challenge for neutrino event generator development work is the wide
range of required expertise, which is cross-cutting along multiple dimensions.
Enhanced support for inter-disciplinary collaboration across theory, experiment, 
and computation, as well as high-energy and nuclear physics is required 
to ensure that neutrino generators {\it i}) 
reflect our best understanding of the underlying scattering physics (and
associated uncertainties), {\it ii}) are responsive to experimental needs and new data sets, and
{\it iii}) adopt best practices for scientific software development and
user support.

Sustained support for event generator development will further be essential in order to ensure that all 
relevant theoretical models are combined consistently in experimental analyses.
Experience and tools from other subfields, notably simulation efforts for
collider physics and simulations of heavy ion collisions, are currently underused and should be explored more
thoroughly by the neutrino generator community.

The content of Sec.~\ref{sec:xsec} was informed by Snowmass White Papers WP\cite{Campbell:2022qmc,Ruso:2022qes} and references therein.

\section{Neutrinoless double-beta decay and other nuclear probes}
\label{sec:nu0betabeta}

The next generation of tonne-scale neutrinoless double-beta ($0\nu\beta\beta$) decay experiments has the opportunity to answer fundamental questions about the nature of neutrino masses, with profound implications for our
understanding of the mechanism by which neutrino mass is generated and of the origin of the matter-antimatter asymmetry in the Universe. 
While the observation of $0\nu\beta\beta$ decay
will imply that neutrinos are Majorana particles, $0\nu\beta\beta$ experiments are sensitive to a variety of lepton-number-violating (LNV) mechanisms, including the standard scenario driven by the exchange of light Majorana neutrinos, low-scale seesaw scenarios with light sterile neutrinos below the electroweak scale, and models of BSM physics with new degrees of freedom at the TeV scale. The observation of $0\nu\beta\beta$ decay could even help solve the flavor puzzle because different classes of flavor models predict different rates of decay.

The interpretation of 
$0\nu\beta\beta$ experiments and, in case of an observation, the 
solution of the ``inverse problem'' of identifying the microscopic mechanism behind a signal demand an ambitious theoretical program to: $a)$ further develop particle-physics models of LNV, including simplified models that go beyond the Majorana neutrino-mass paradigm, 
and test them against the results of current and future $0 \nu \beta \beta$ 
experiments, the Large Hadron Collider (LHC), and astrophysics and cosmology;
$b)$  compute $0\nu\beta\beta$ rates with minimal model dependence and quantifiable theoretical uncertainties by advancing progress in particle and nuclear effective field theories (EFTs), lattice quantum chromodynamics (QCD), and nuclear few- and many-body \textit{ab initio} methods. 

\subsection{Bridging particle and nuclear physics with EFTs}

\begin{figure}[!t]
    \centering
    \includegraphics[width=\textwidth]{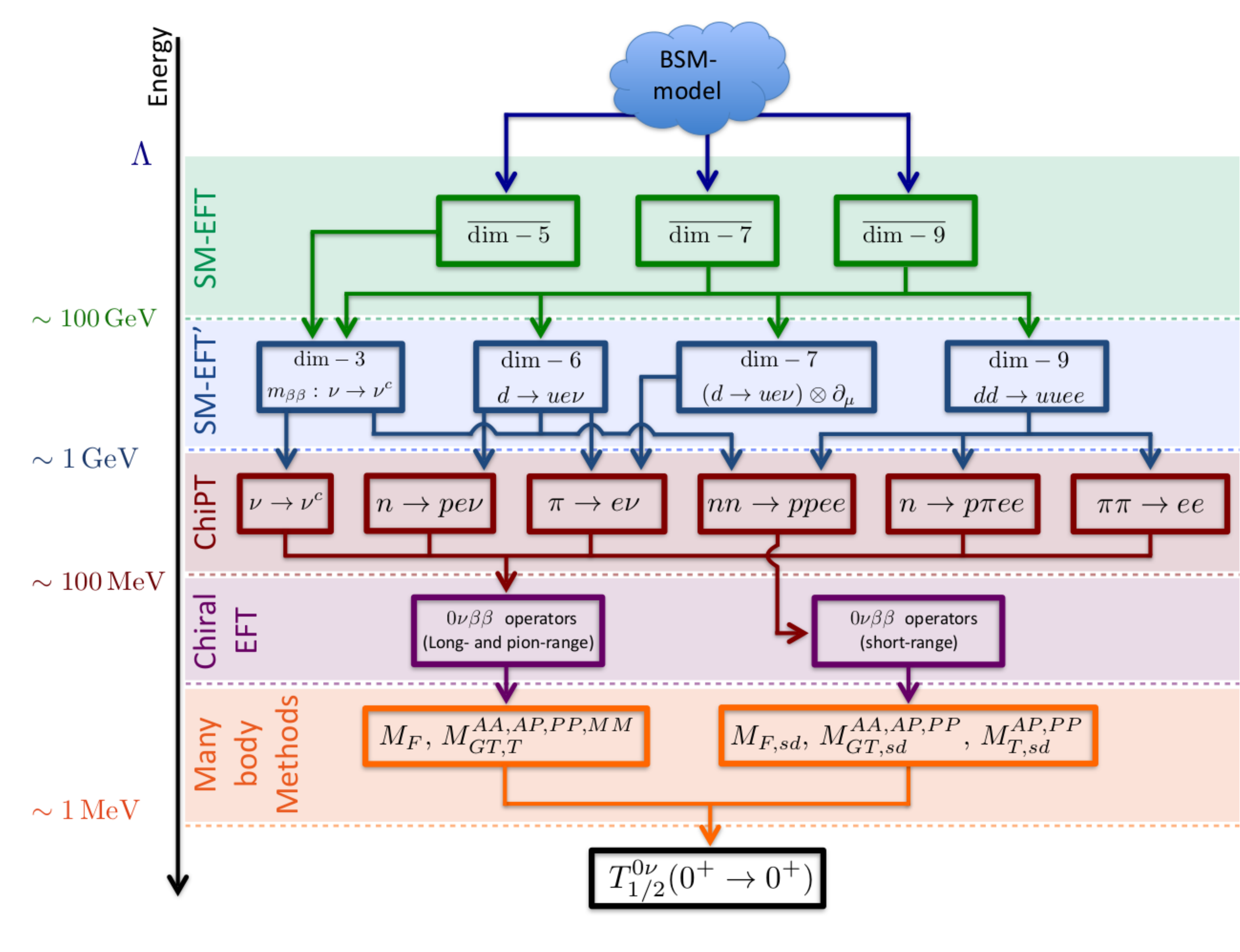}
    \caption{The tower of EFTs for $0\nu\beta\beta$ decay. At the electroweak scale, LNV operators are described by operators of odd dimension in the SMEFT. Heavy SM degrees of freedom can be integrated out by matching SMEFT onto LEFT (denoted as SMEFT$^\prime$ in the figure). Quark-level operators are then matched onto hadronic operators. The construction of hadronic operators is performed in $\chi$PT and chiral EFT, while the determination of the low-energy couplings requires non-perturbative techniques, such as lattice QCD. The $0\nu\beta\beta$ transition operators constructed in chiral EFT are then evaluated with nuclear many-body methods.
    Figure is adapted from Ref.~\cite{Cirigliano:2018yza}.}
    \label{fig:landscape}
\end{figure}

The steps needed to achieve b) can be organized in terms of several EFTs, as schematically depicted in Fig.\ \ref{fig:landscape}. If one assumes LNV arises at a scale, $\Lambda$, well above the electroweak scale, the LNV interactions can be described by an EFT consisting of the SM extended by higher-dimensional operators built from SM fields, the so-called SMEFT (the top layer of Fig.\ \ref{fig:landscape}). To assess their impact on $0\nu\beta\beta$ rates, the LNV operators need to be evolved to lower energies. In doing so one first encounters the electroweak scale, where the heavy SM particles ($W,\, Z,\, t,\, h$) are integrated out, leading to a second EFT often referred to as the LEFT (blue layer). After evolving this theory to the QCD scale its quark-level operators can be matched onto interactions in chiral EFT in terms of nucleons, pions, and leptons (the purple layer in the figure). This matching involves low-energy constants (LECs) which parametrize non-perturbative matrix elements that can be obtained from lattice QCD calculations. Finally, many-body calculations (orange layer) are needed to go from the LNV nucleon-nucleon potential, derived from chiral EFT, to the decay rates for the nuclei of interest. 

Below, we briefly discuss the most important steps as well as the interplay with collider observables. Simplified models of LNV at the TeV scale provide useful schemes for investigating correlated signals of LNV across multiple types of experiments, spanning wide energy and distance scales. For certain sources of  LNV, LHC experiments have the potential to surpass constraints from $0\nu\beta\beta$ experiments and  to explore complementary regions of parameter space. We encourage the LHC collaborations  to investigate simplified models for TeV-scale LNV and to make projections for the high-luminosity LHC (HL-LHC). LHC searches will help exclude or discover alternative TeV-scale interpretations of a $0\nu\beta\beta$ signal, and, through discovery, 
help falsify high-scale models of leptogenesis. Finally, both full and simplified BSM models can be matched to hadronic EFTs, which allow for a systematic expansion of the $0\nu\beta\beta$ rates.
     
\subsection{Lattice-QCD input for neutrinoless double-beta  decay} 

The EFTs that systematically classify LNV in the few-nucleon sector need to be complemented with values for low-energy constants (LECs).  The lack of experimental input makes lattice QCD, which computes the relevant matrix elements directly, the only way to determine these LECs.  To systematically and reliably match lattice QCD matrix elements to EFT LECs, we require a) a targeted computational campaign, powered by high-performance computing, to obtain precise two-nucleon spectroscopy and matrix elements, and b) theoretical developments to clarify the path from lattice-QCD output to physical matrix elements in the two- and higher-nucleon sectors, within various high-scale models of LNV.

\subsection{Nuclear structure and the computation of 
matrix elements} 

With the systematic EFTs constrained by lattice QCD, uncertainties in nuclear matrix elements (NMEs) of relevance to experimental isotopes can be realistically assessed and reduced. 
Under the auspices of the Topical Collaboration on Nuclear Theory for Double-Beta Decay and Fundamental Symmetries, researchers began to shift from phenomenological approaches to \textit{ab initio} methods for the calculation of $0\nu\beta\beta$ NMEs, with the first wave of \textit{ab initio} results for $^{48}$Ca, $^{76}$Ge,
and $^{82}$Se appearing in the last few years.
Work is already underway to validate and improve the approximations used in these calculations. 
The computation of next-generation NMEs for $0\nu\beta\beta$ candidate nuclei will require considerable amounts of computing time as well as investments in the
development of many-body codes to ensure that the allocated time is used efficiently. Parallel advances in nuclear EFTs and lattice QCD are required to guarantee that the nuclear interactions and transition operators are constructed at the same order and in the same regularization scheme. Finally, the community needs to address deep questions about the 
implementation of EFT interactions and operators in traditional many-body methods. 
%
Section~\ref{sec:nu0betabeta} was informed by the solicited Snowmass White Papers~WP\cite{Cirigliano:2022oqy,Gehrlein:2022nss}.

\section{Outlook}

Neutrinos have revealed that our understanding of fundamental particle physics is incomplete. Nonzero neutrino masses imply the existence of new particles and interactions. We know very little about these new particles. They can be right-handed neutrinos, new Higgs fields, or collections of new, charged particles that live above the weak scale, to name a few. A diverse experimental neutrino program is underway, aimed at exploring the new physics revealed in the neutrino sector. It includes very intense neutrino beams, very large, finely instrumented detectors, very large, ultra-clean detectors to search for neutrinoless double-beta decay, and novel detectors for precision measurements of beta-decay. In the next decade, a deluge of neutrino-related data is expected. A concurrent theoretical and phenomenological neutrino physics effort is required, on all fronts.

The theoretical neutrino effort is both very broad in terms of the tools required and rather focused  when it comes to individual physics challenges. In summary, neutrino theory requires a broad set of tools in order to attack a unique set of physics problems. `Neutrino theory' deals with energy scales as low as sub-eV and as high as an EeV. It includes astrophysics, cosmology, particle physics phenomenology and model building, nuclear physics, and lattice gauge theory, entwining different areas of physics and astronomy.   

Front and center is the issue of fully exploiting the US and Worldwide investment made in DUNE and the Long-Baseline Neutrino Facilty (LBNF). Particle physics phenomenology and model building will play a vital role in defining physics questions for the near and far detector complexes and interpreting the experimental results. These will help define future directions for neutrino physics. More urgent, perhaps, is the need to understand the physics of neutrino--scattering across energy scales that range from the quasi-elastic regime to shallow and deep inelastic scattering. Support for theoretical efforts on neutrino scattering at the interface of high-energy and nuclear physics is critical for achieving reliable cross-section predictions across the range of energies relevant to current and planned neutrino experimental programs.
Programs that support the collaboration of theorists working on specific topics across nuclear and high-energy physics are necessary to enable progress in this area.  Increased support is also needed to implement models into and maintain neutrino event generators.

\bibliographystyle{JHEP}

\providecommand{\href}[2]{#2}\begingroup\raggedright\endgroup

\end{document}